%Paper: hep-ph/9206209
%From: "Jiang Liu" <JIANGLIU@penndrls.upenn.edu>
%Date: Thu, 4 Jun 92 13:48:26 EST

\input phyzzx
\PHYSREV
\hbox to 6.5truein{\hfill UPR-0495T}
\hbox to 6.5truein{\hfill June 1992}
\title{STANDARD MODEL CONTRIBUTIONS TO THE NEUTRINO  INDEX OF REFRACTION
       IN THE EARLY UNIVERSE}
\author{Paul Langacker and Jiang Liu}
\address{Department of Physics, University of Pennsylvania, Philadelphia,
         PA 19104}
\abstract
With the standard electroweak interactions, the lowest-order
coherent forward scattering amplitudes of
neutrinos in a CP symmetric medium (such as the early
universe) are zero, and the index of refraction
of a propagating neutrino can
only arise from the expansion of  gauge
boson propagators, from radiative corrections, and from new physics
interactions.
Motivated by nucleosynthesis constraints on a possible sterile
neutrino (suggested by the solar neutrino deficit and a possible
$17\ keV$ neutrino),
we calculate the standard model
contributions to the neutrino index of refraction in the early universe,
focusing on the period when
the temperature  was of the order of a few $MeV$.
We find sizable radiative corrections to the tree level result
obtained by the expansion of the gauge boson propagator.
For  $\nu_e+e(\bar{e})\to \nu_e+e(\bar{e})$ the leading
log correction is about $+10\%$, while for
$\nu_e+\nu_e(\bar{\nu}_e)\to \nu_e+\nu_e(\bar{\nu}_e)$ the correction
is about $+20\%$.   Depending on the family mixing (if any),
effects from
different family scattering  can be dominated by radiative
corrections.
The result for $\nu+\gamma\to\nu+\gamma$ is
zero at one-loop level, even if neutrinos are massive.
The  cancellation of
infrared divergence in a coherent process is also discussed.

PACS\# 13.10.+q, 13.15.-f
\endpage

\chapter{Introduction}

Neutrinos propagating in a medium could behave differently from
in  vacuum.  The effect of the medium on the neutrino can
conveniently be described by
the index
of refraction, $n$.  Under certain conditions,  matter effects could be
so enormous that neutrino properties are modified drastically.
The best known example is  the Mikheyev-Smirnov-Wolfenstein (MSW)
mechanism\Ref\MSW{L. Wolfenstein, Phys. Rev. D17, 2367 (1978);
         {\it {ibid}}, D20, 2634 (1979);
         S. P. Mikheyev and A. Yu. Smirnov,  Yad. Fiz. 42, 1441 (1985)
         [Sov. J. Nucl. Phys. 42, 913 (1985)].}
for the solar neutrino
deficit\rlap,\REFS\solar{J. K. Rowley, B. T. Cleveland and R. Davis, Jr.,
         in J. Schneps {\it{et al., Proc. XIIth Intl. Conf. Neutrino
         Phys. \& Astrophys. (NEUTRINO '88)}} (Singapore: World
         Scientific, 1989) 518.}\REFSCON\komi{
         K. S. Hirata et al., Phys. Rev. Lett. 63, 16 (1989),
         {\it {ibid}}, 65, 1297 (1990), {\it{ibid}}, 66, 9 (1991).}
          \refsend
in which the index of refraction of
 electron neutrinos propagating inside the sun could
induce
a sufficiently large effective mixing angle for them to oscillate
into other types of neutrinos efficiently.

Neutrino oscillations could also be cosmologically
important\rlap.\Ref\cosmol{P. Langacker, B. Sathiapalan, and
         G. Steigman, Nucl. Phys. B266, 669 (1986);
         M. Yu Khlopov and S. T. Petcov, Phys. Lett. 99B, 117 (1981),
         100B, 529E (1981).}
One particularly interesting
example\rlap,\Ref\sterile{P. Langacker,  University
of Pennsylvania preprint
         UPR 0401T (1989) (unpublished).}
 pointed out by one of us (P.L.)
some time ago, would be oscillations between a standard neutrino and
a light sterile $SU(2)_L$ singlet neutrino, $\nu_s$.
If certain conditions are satisfied prior to neutrino decoupling
 such oscillations
can bring $\nu_s$ into equilibrium and thus alter the expansion
rate of the universe.
In particular, for a wide range of masses and mixings one would have
$N_{\nu}=4$ neutrino species in equilibrium, in violation of the bound
$N_{\nu}<3.3$ $(95\%$ CL) from the observed abundance of
$^4He$ and other light elements\rlap.\Ref\synthesis{
     K. A. Olive et. al., Phys. Lett. B236, 454 (1990);
     T. P. Walker et. al., Astrop. J. (to be published).}
Subsequent more detailed studies\REFS\Barie{
     R. Barbieri and A. Dolgov, Phys. Lett. B237, 440 (1990),
     Nucl. Phys. B349, 742 (1991).}\REFSCON\enqv{
     K. Enqvist, K. Kainulainen, and J. Maalampi, Phys. Lett. B244,
     186 (1990), B249, 531 (1990);
     K. Enqvist, K. Kainulainen, and M. Thomson, NORDITA preprint\break
     91/71 P.}\REFSCON\Mck{
     M. J. Thomson, Bruce H. J. Mckellar, Phys. Lett. B259, 113 (1991);
     University of Melbourne prepint, UM-P-90/44; UM-P-90/37
     (unpublished).}
     \refsend
have
 reached similar conclusions.

Theoretically, the possibility of having a standard neutrino
oscillate to $\nu_s$ has long been of interest.  Many theoretical
models based on the compactification of string theories and
grand unification predict the existence of light sterile neutrinos.
A recent study\Ref\simpson{D. O. Caldwell and P. Langacker,
    Phys. Rev. D44, 823 (1991).}
shows that if the controversial 17-$keV$ neutrino exists,
laboratory, astrophysical, and cosmological bounds, unless
significantly weakened,
requires $m_{\nu_{\tau}}=17\ keV$ and $m_{\nu_{\mu}}\ge m_{\nu_{\tau}}$.
Thus, the MSW solution to the solar neutrino deficit would require
$\nu_e-\nu_s$ oscillations.  It turns out that the mass and mixing
parameters required for the MSW solution are close to the boundary that
is excluded by primordial nucleosynthesis.  Moreover, the exact
location of that boundary depends sensitively on the index of refraction
and other details.

An accurate determination of the neutrino index of refraction
is therefore  an important step in the study of neutrino
oscillations and their applications to solar neutrinos and cosmology.
  In  the zero temperature limit,
this can be done easily by a direct calculation of neutrino
forward scattering amplitudes.  The situation is more
complicated if the thermal motion of  scatterers
becomes important.  For that,
lowest order calculations\Ref\index{
    D. Notzold and G. Raffelt, Nucl. Phys. B307, 924 (1988);
    P. B. Pal and T. N. Pham, Phys. Rev. D40, 714 (1989);
    J. F. Nieves, Phys. Rev. D40, 866 (1989);
    K. Enqvist, K. Kainulainen and J. Maalampi, Nucl. Phys. B349,
    754 (1991).}
have appealed to a more complicated method derived
from  finite
 temperature field
theory\rlap.\Ref\ftft{
    L. Dolan and R. Jackiw, Phys. Rev. D9, 3320 (1974);
    S. Weinberg, {\it{Gravitation and Cosmology: Principles and
    Applications of the General Theory of Relativity.}} New York,
    Wiley (1979);
    H. A. Weldon, Phys. Rev. D26, 2789 (1982).}

In this article, we wish to study the standard electroweak
contributions to the neutrino index of refraction in the early universe.
The epoch of interest is radiation dominated and has a temperature $T$
of the order of a few $MeV$.  At that time,
according to the standard Big Bang cosmology,
the universe consisted mainly of photons
$\gamma$, electrons, positrons, and three types of neutrinos
and  anti-neutrinos.  Motivated by the smallness of the baryon
asymmetry, we will assume, for simplicity, that
in that era the universe was CP symmetric\rlap.\Ref\aaa{
  Assuming the lepton asymmetry is of the same order of magnitude as
  the baryon asymmetry of the order $10^{-9}-10^{-10}$,\splitout
  the effect of CP violation of the universe in the era is not important
  (Ref. \Barie ).}
As a consequence, there
were equal amount of particles and antiparticles, with densities
given by
$$\eqalign{&N_{\gamma}={2\zeta (3)\over \pi^2}T^3,\cr
&N_e=N_{\bar{e}}={3\over 4}N_{\gamma},\cr
&N_{\nu_e,\nu_{\mu},\nu_{\tau}}=N_{\bar{\nu}_e,\bar{\nu}_{\mu},
\bar{\nu}_{\tau}}={3\over 8}N_{\gamma},\cr}\eqno(1.1)$$
where $\zeta (3)=\sum_{n=1}^{\infty}n^{-3}\approx 1.202$.

In terms of the standard electroweak interactions,
the CP symmetry of the medium implies that the  lowest order
terms in  neutrino coherent forward scattering amplitudes cancel.
This follows because the  amplitude of, say,
neutrino-electron scattering is equal in amplitude but opposite in sign
 to that of
neutrino-positron scattering.
As a result, the neutrino index of refraction is zero to  lowest order.
Indeed, previous calculations ( Ref. \index)  have shown that
the only nonzero result arises from the expansion of the $W$ and  $Z$
boson propagators, and hence is suppressed  by
a very small factor of  order $q^2/M_W^2$, where $q$ is the momentum
carried by the $W$ or $Z$.    In the absence of family mixing, even such
a tiny result vanishes for scatterings that take place between
different families.

The smallness of the tree level result  calls for an
investigation of its radiative corrections.
In some places,
particularly in resonance regions,  theoretical solutions
are   sensitive to the  value
of neutrino  index of refraction.
Loop effects at zero termperature\Ref\radiative{
     L. M. Sehgal, Phys. Lett. B162, 370 (1985);
     F. J. Botella, C.-S. Lim, and W. J. Marciano, Phys. Rev. D35, 896
     (1987);  R. Horvat and K. Pisk, Nuov. Cim. 102A, 1247 (1989).}
have been discussed before.  One finds that they
introduce new interactions proportional to
$(\alpha/\pi)(m_{\ell}^2/M_W^2)\ln(m_{\ell}^2/M_W^2) (\ell=e,\mu,\tau)$
that break the universality of the neutral-current.
 In this paper, this effort will
be carried over to the finite temperature case.
Due to the aforementioned cancellation, radiative
corrections now take a completely different form.

The remainder of this paper is organized as follows.  In the next section
we reformulate the lowest order calculation by introducing a method
that uses  standard field theory rather than the finite temperature
field theory.  Within this framework,  higher order
calculations are simplified to a standard calculation of
 radiative corrections
to  neutrino forward scattering amplitudes.  In section
3, we present the leading log result for the $
\nu_e+e(\bar{e})\to \nu_e+e(\bar{e})$ scattering.  There, we
begin with a discussion of the compatibility of coherent conditions
with the cancellation of infrared divergence.  Neutrino-neutrino
scattering and its radiative corrections are discussed in section 4.
Scattering between different generations of neutrinos and leptons
is discussed in section 5.
Neutrino-photon scattering is discussed in section 6.
The imaginary part of  neutrino index of refraction is given in
section 7. Our results are summarized in the final section and
some technical details are presented in the appendix.
The application of our results to cases of cosmological and
astrophysical interest will be discussed elsewhere\rlap.\Ref\PJ{
P. Langacker and J.  Liu (to be published).}

\chapter{General Formula}

At zero temperature the index of refraction, $n$, is given by
\Ref\fermi{E. Fermi, {\it{Nuclear Physics}}, pp201,
       the University of Chicago press, (1949);  L. L. Foldy, Phys. Rev.
       67, 107 (1945);  M. Lax, Rev. Mod. Phys. 23, 287 (1951);
       M. L. Goldberger and K. M. Watson, {\it{Collision Theory}},
       P768, John Wiley \& Sons, Inc. (1964); see also the first article
       in Ref. 1.  Eq. (2.1) holds only if $\vert n-1\vert\ll 1$.
       In general,  the  formula for the index of refraction is
       $n^2-1=4\pi Nf(0)/p_0^2$.}
$$ n-1={2\pi\over p_0^2}Nf(0),\eqno(2.1)$$
where $p_0$ is the energy of the propagating
 neutrino, $N$ is the number of
scatterers per unit volume, and $f(0)$ is the neutrino forward scattering
amplitude.  In the MSW formula, for example,
$f(0)$ arises from $\nu_e-e$
scattering via the charged-current interaction, where\Ref\levi{
    P. Langacker, J. P. Leveille, and J. Sheiman, Phys. Rev. D27,
    1228 (1983).}
in the rest frame of the electron
$f(0)=G_Fp_0/\sqrt 2\pi$, and hence $n=1+\sqrt 2NG_F/p_0$.

The fundamental formula  Eq. (2.1) can be
generalized to situations in which
 the medium has a finite temperature $T$.
With the usual assumption that
the introduction of a finite temperature
does not spoil the coherent
condition\rlap,\Ref\spoil{
  This is a nontrivial assumption.  The detailed study of this question
    lies beyond the scope of the present paper.}
the only modification is to take the
thermal average
$$n-1={2\pi\over p_0^2}
\langle Nf(0)\rangle={2\pi\over p_0^2}
\langle N\rangle \langle f(0)\rangle,\eqno(2.2)$$
where the last step follows from the usual assumption that scatterers
are not correlated.
More explicitly
$$\eqalign{
&\langle N\rangle=\int{d^3\vec{q}\over (2\pi)^3} n^*(q_0, T),\cr
&\langle f(0)\rangle
=\langle N\rangle^{-1}\int{d^3\vec{q}\over (2\pi)^3}n^*(q_0,T)f(0),\cr}
\eqno(2.3)$$
where
$$n^*(q_0,T)={g^*\over e^{(\mu +q_0)/kT}\pm 1}   \eqno(2.4)$$
is the  occupation number density of the scatterer,
$\mu$ is the chemical potential,
and $g^*$ is the number of spin degrees of freedom.
Within this simple scheme, physical quantities such as
the forward scattering amplitude are calculated
  in the standard way (at zero temperature), and the final result is
obtained by
taking the thermal average according to Eq. (2.2) at the end.

This simple method\Ref\equival{  The equivalence between the present
  method and the finite temperature method can be shown diagrammatically
  in perturbation theory.}
 makes the lowest order calculation of neutrino
index of refraction trivial.  As an illustration,  consider
the scattering $\nu_e(p)+e(p')\to \nu_e(p)+e(p') $.  The tree
level Feynman graphs are shown in Fig. 1. Neglecting the electron
and the neutrino mass,
the scattering matrix element
is  given by
$$\eqalign{M_0(\nu_ee\to\nu_ee)=&-\Bigl({ig\over\sqrt 2}\Bigr)^2
    {-i\over (p-p')^2-M_W^2}[\bar{u}_e(p')\gamma_{\alpha}Lu_{\nu_e}(p)]
                            [\bar{u}_{\nu_e}(p)\gamma^{\alpha}Lu_e(p')]
\cr
&-\Bigl({ig\over 2\cos\theta_W}\Bigr)^2{-i\over 0-M_Z^2}
      [\bar{u}_e(p')\gamma_{\alpha}(L-2\sin^2\theta_W)u_e(p')]
      [\bar{u}_{\nu_e}(p)\gamma^{\alpha}Lu_{\nu_e}(p)],\cr}
\eqno(2.6)$$
where  $\theta_W$ is the weak mixing angle, and
$u_e$ and $u_{\nu_e}$ are the electron and neutrino
spinors, respectively.  The first term in Eq. (2.6) arises
from the $ u$-channel scattering  (Fig. 1a), and the second
comes from the $t$-channel (Fig. 1b), where
due to the  coherent condition the momentum transfer must be zero.
The $u$-channel scattering matrix element has an additional minus sign,
which can be understood as a consequence of  fermion line crossing.
One finds from Eq. (2.6) that the {\it{coherent}}
 forward scattering amplitude is
$$f(0)={\sqrt 2 G_F p.p'\over 2\pi p_0'}
\Bigl[{1+4\sin^2\theta_W
\over 2}-{2p.p'\over M_W^2}\Bigr]+...,\eqno(2.7)$$
where the first term in the parentheses is the same as that
obtained from a local $V-A$ interaction (i.e., $(p-p')=0$),
the second arises from the expansion of $W$ propagator, and
the ellipses refer to the higher order terms.
It then follows from Eqs.(2.2) and (2.7) that
$$(n-1)_{\nu_ee\to\nu_ee}={\sqrt 2G_F\over p_0}
\Bigl[{1+4\sin^2\theta_W\over 2}N_e-{8\over 3M_W^2}p_0\langle p_0'\rangle
N_e\Bigr],\eqno(2.8)$$
where $N_e$ is the electron number density at temperature $T$
and $\langle p_0'\rangle$ is the average electron energy.  In reaching
Eq. (2.8), we have assumed that the thermal motion of the electrons is
isotropic so that $\langle \vec{p'}\rangle=0$.

The matrix element for
$\nu_e(p)+\bar{e}(p')\to\nu_e(p)+\bar{e}(p')$ (Fig. 2)
can be obtained from Eq. (2.6)
by changing $u_e\to v_e=u_e^c$, $p'\to -p'$, and introducing an over all
minus sign (due to the anticommutation  of fermion fields).
One finds
$$(n-1)_{\nu_e\bar{e}\to\nu_e\bar{e}}=
{\sqrt 2 G_F\over p_0}\Bigl[-{1+4\sin^2\theta_W
\over 2}N_{\bar{e}}-{8\over 3M_W^2}
      p_0\langle p_0'\rangle N_{\bar{e}}\Bigr].\eqno(2.9).$$
The sum of Eqs. (2.8) and (2.9) gives
$$(n-1)_{\nu_ee(\bar{e})\to\nu_ee(\bar{e})}
={\sqrt 2G_F\over p_0}\Bigl[{1+4\sin^2\theta_W\over 2}(N_e-N_{\bar{e}})
-{8p_0\over 3M_W^2}(\langle
p_0'\rangle_eN_e+\langle p_0'\rangle_{\bar{e}}N_{\bar{e}})\Bigr].
\eqno(2.10)$$
This is precisely the result  obtained previously by  the
finite temperature field theory method (Ref. \index ),
in which  one
calculates one-loop contributions to neutrino self-energy in a
background field.  The present method is considerably simpler, since it
only involves a standard tree level graph calculation\rlap.\Ref\kim{
     A similar
     method has previously been proposed in the context of neutrino decay
     in matter:  C. Giunti, C. W. Kim, and W. P. Lam, Phys. Rev. D43,
     164 (1991).}

Our approach also makes the underlying physics more intuitive.
That the first term in Eq. (2.10) is given by the difference
of the electron and  positron densities is a consequence of
the CP transformation property of vector currents
$$\bar{u}_e\gamma_{\alpha}u_e\to -\bar{v}_e\gamma_{\alpha}v_e
=-\bar{u}_e^c\gamma_{\alpha}u_e^c=-\bar{u}_e\gamma_{\alpha}u_e.
\eqno(2.11)$$
  Diagrammatically,
from Fig.(1) and Fig.(2) the
$t$-channel coherent scattering
amplitudes are equal in size
but  opposite in sign, and
hence cancel completely.  The same would also be
true if the $W$'s in Fig. (1a) and  Fig. (2a)
were not carrying any momenta.  In the neutrino-electron
scattering the momentum carried by the $W$ is $p'-p$, whereas in the
neutrino-positron scattering it changes to $p'+p$.
This relative sign
difference compensates the over all sign difference discussed
in Eq. (2.11), so
that the second term of Eq. (2.10) becomes a sum rather than a
difference.   Finally, contributions from the longitudinal part of the
gauge boson propagators are always negligible if the external
fermion masses are small.  These terms have the same structure as that
in Eq. (2.10) times a multiplicative factor given by a ratio like
$m_e^2/M_W^2$.

Similar observations
can also be made for loop graphs that generate
radiative corrections to the tree level result.  Basically,
all $t$-channel one-particle-reducible graphs will cancel, and hence
we will omit them from now on.  This includes all  zero temperature
correction results discussed in Ref. \radiative .
  Only the
$s$- and $u$-channel one-particle-reducible diagrams and the box diagrams
may  contribute to the radiative corrections.

\bigskip

\chapter{One-loop Corrections to the
$\nu_e+e(\bar{e})\to \nu_e+e(\bar{e})$ Forward Scattering Amplitude}

The order
$\alpha/\pi$ correction results
are known\rlap.\REFS\ueda{
    P. Salomonson and Y. Ueda, Phys. Rev. D11, 2606
    (1975).}\REFSCON\passarino{G. Passarino and M. Veltman, Nucl. Phys.
    B160, 151 (1979).}
    \REFSCON\Green{M. Green and M. Veltman, Nucl. Phys. B169,
    137 (1980), (E: B175, 547 (1980)).}\REFSCON\Aoki{K-I. Aoki,
    Z. Hioki, R. Kawabe, M. Konoma and T. Muta, Prog. Theor. Phys.
    65, 1001 (1981); K-I. Aoki, and Z. Hioki, Prog. Theor. Phys.
    66, 1165 (1982).}
    \REFSCON\SSM{
    S. Sarantakos, A. Sirlin, and W. J. Marciano, Nucl. Phys. B217,
    84 (1983).}\refsend
 However,
they have the same structure as the lowest order weak interaction,
and thus  the sum of the coherent amplitudes for
 neutrino-electron and  neutrino-positron
scattering  cancels.

Terms which may survive the cancellation
must depend on $ (p\pm p')^2$ where the two signs refer to
neutrino-electron
and  neutrino-positron scattering, respectively.
 Comparing with the tree level
result $(\sim G_F(p\pm p')^2/M_W^2)$ arising from the expansion of
the gauge boson propagator,
loop corrections are
of the order $(\alpha/4\pi)
G_F(p\pm p')^2/M_W^2$ times  a yet to
be determined factor.  Normally, this factor contains large log
terms plus some constant terms of  order unity.
Since the ratio $\alpha/4\pi$ is already  small,
in the absence of logarithmic enhancement radiative corrections would be
negligible.  Thus,
in what follows we
will only keep those leading log terms.

Explicit calculations (see below) show that there are only
two scalar functions arising from loop integrals
 that  have a large log
$$\eqalign{&I_1(m_1,m_2,M;q^2)
\equiv {q^2\over M^2}\int_0^1dx x(x-1)
\ln {m_1^2+x(m_2^2-m_1^2)+x(x-1)q^2\over M^2},\cr
&I_2(m_1,m_2,M;q^2)\equiv
\int_0^1dx{m_1^2+x(m_2^2-m_1^2)\over M^2}\ln{
           m_1^2+x(m_2^2-m_1^2)+x(x-1)q^2\over M^2},\cr}
           \eqno(3.1)$$
where $m_{1,2}$ are two small masses, one for a neutrino and another for
a charged lepton.  In Eq. (3.1),
 $M$ is either $M_W$ or
$M_Z$, and $q=p\pm p'$ is the momentum of the virtual $W$ or $Z$.
The logarithmic enhancement arises because
of the hierarchy:
$$ m_{1,2}^2, \vert q^2 \vert \ll M^2,\eqno(3.2a)$$
and hence $\ln(M^2/m_{1,2}^2), \ln(M^2/\vert p\pm p'\vert^2)\gg 1$.
Consistent with our leading log approximation,
terms proportional to $I_2$ can  be neglected, because
the difference, which one must take in the end when summing over the
neutrino-electron and the neutrino-positron scattering amplitudes,
$$\eqalign{&I_2(m_1,m_2,M;(p+p')^2)-I_2(m_1,m_2,M;(p-p')^2)\cr
&=
\int^1_0dx{m_1^2+x(m_2^2-m_1^2)\over M^2}\ln{m_1^2+x(m_2^2-m_1^2)
+x(x-1)(p+p')^2\over
m_1^2+x(m_2^2-m_1^2)+x(x-1)(p-p')^2}\cr}\eqno(3.2b)$$
does not have a large log.   Effectively, this implies that we can
neglect all  small masses whenever possible
in our calculation.  A welcome
consequence of such a simplification is that  all graphs
that contain scalars (in the Feynman gauge) can be neglected.
As long as we  keep only the leading log term,
the result will be gauge invariant.
We will assume that neutrino masses are
completely negligible.

\section{Cancellation of Infrared Divergences}

A  question which one might encounter in the study of
radiative corrections  is
how do the infrared divergences\Ref\block{
    F. Block and A. Nordsieck, Phys. Rev. 52, 54 (1937).}
 cancel in a coherent process?
In normal (incoherent) cases, the complete cancellation of
infrared divergence is well understood\rlap.\Ref\infrared{
      See for example: D. R. Yennie, S. C. Frautschi, and H. Suura,
      Ann. Phys. 13, 379 (1961); S. Weinberg, Phys. Rev. 140, B516
      (1965).}
Although some individual Feynman graphs generate infrared divergent
terms, they cancel when  summed  with
the bremmstrahlung diagrams, which are also infrared divergent
by themselves.

In a coherent process, however, bremsstrahlungs cannot be included.
Thus, if a process is to remain as coherent at higher order, the
cancellation of infrared divergences must take place among  themselves.
In  the MSW formalism,  the constraint becomes even more restrictive.
There, the neutral- and the charged-current interaction must be
infrared finite seperately.

The cancellation of infrared divergences
for a coherent interaction
 has not yet
received much attention,
partly because there are not many places in which the physical
process can be coherent.  Marciano and Sirlin have investigated
this question\Ref\atomic{
   W. J. Marciano and A. Sirlin, Phys. Rev. D27, 552 (1983),
   {\it{ibid}}, D29, 75 (1984).}
in the context of atomic parity violation.
By explicit calculations, they
show that the required cancellation
does occur at one-loop level.  In the following, we will show
that this happens to all order in  perturbation theory.

Consider an electron scattering with an infrared finite
but otherwise arbitrary potential  and its  QED
corrections (Fig.3).  Suppose the
forward scattering amplitude of the interaction is
$$M_0(p,p)=\bar{u}_e(p)\Gamma_0(p,p)u_e(p),\eqno(3.3)$$
where $\Gamma_0(p,p)$ is the interaction vertex,
which is assumed to be renormalizable so that it is also free
from ultra-violet divergences.

At one-loop level, QED corrections introduce infrared divergences into
some of the Feynman graphs.
To isolate them, we recall that infrared
divergences arise because photons are massless, and thus a slight
acceleration of an external charged particle could result in
an emission of an infinite number of soft photons.  This implies
that only those Feynman graphs in which a virtual photon line has
both its ends  attached to an external line (Fig. 3b to 3d) are infrared
divergent (Ref. \infrared ).  The infrared singularity is generated by
the photon pole
$${-i\over k^2-\lambda^2+i\epsilon}=P.V. {-i\over k^2-\lambda^2}-
\pi\delta(k^2-\lambda^2),\eqno(3.4)$$
where $\lambda$ is a ficticious photon mass.
The one-loop QED corrections can now be written as
$$\delta M_1(p,p)=\delta M_1^{(b)}(p,p)+\delta M_1^{(c)}(p,p)
                 +\delta M_1^{(d)}(p,p)+(infrared\ finite\ terms),
\eqno(3.5)$$
where $\delta M_1^{(b)}(p,p)$ is the one-loop vertex correction (Fig. 3b)
$$\eqalign{&
\delta M_1^{(b)}(p,p)=\cr
&-ie^2\int {d^4k\over (2\pi)^4}\bar{u}_e(p)
\gamma_{\alpha}{1\over p.\gamma+k.\gamma-m_e}\Gamma_0(p+k,p+k)
{1\over p.\gamma +k.\gamma-m_e}\gamma^{\alpha}u_e(p){1\over k^2-\lambda^2
+i\epsilon},\cr}\eqno(3.6)$$
and $\delta M_1^{(c)}(p,p)$ and $\delta M_1^{(d)}(p,p)$ are the
one-loop wave function renormalization contributions.

In the on-shell subtraction
scheme, one finds  easily
$$\delta M_1^{(c)}(p,p)+\delta M_1^{(d)}(p,p)=-M_0{\alpha\over 2\pi}
\ln{\lambda^2\over m_e^2}+(infrared\ finite\ terms).\eqno(3.7)$$
The infrared divergence
in the vertex correction can be separated out by considering the
photon pole (Eq.(3.4)) contribution to the
matrix element in the infrared region, $k\to 0$, where by  virtue
of the finiteness of $M_0(p,p)$ one finds
$\lim_{k\to 0}\bar{u}_e(p)\Gamma_0(p+k,p+k)u_e(p)=M_0(p,p)$ and thus
$$\eqalign{\delta M_1^{(b)}(p,p)\vert_{0\le k_0\le \omega_0}&=
-e^2M_0(p,p)\int {d^4k\over (2\pi)^4}{m_e^2\over (p.k)^2} \pi
\delta(k^2-\lambda^2)+(infrared\ finite\ terms)\cr
&=-M_0(p,p){\alpha\over 2\pi}\ln{\omega_0^2\over \lambda^2}
+(infrared\ finite\ terms),\cr}
\eqno(3.8)$$
where $\omega_0$ is an infrared cut-off often determined by the
resolution of the experimental apparatus.

It is now evident that the infrared divergence in the wave function
renormalization cancels the infrared divergence in the vertex correction.
The net result is infrared finite
$$M_0(p,p)+\sum_{i=a,b,...}\delta M_1^{(i)}(p,p)=
M_0(p,p)\Bigl(1-{\alpha\over 2\pi}\ln {\omega_0^2\over m_e^2}\Bigr)
+...,\eqno(3.9)$$
where the ellipses represent the other infrared finite terms.

The factorization in the vertex correction (Eq. (3.8)) is crucial in
reaching our final result.  Now, by an elementary induction
and by employing the same method presented above,
one can
readily show that at  N-loop level, infrared divergences from
 the effective vertex (infrared finite at
 N-1 loop level) always
 cancel those from wave function renormalization.
Thus,
the forward scattering amplitude is
infrared finite to all orders.
This is to be expected,
for
if the electron forward scattering amplitude were not infrared finite,
it would have implied that the electric charge is not a well defined
quantity.

\section{QED Correction}

It is convenient to separate out the conventional photonic correction
from the weak interaction corrections.  To do so, we employ the standard
method\Ref\decomp{A. Sirlin, Rev. Mod. Phys. 50, 573 (1978).} to
decompose the photon propagator
$${-i\over k^2}={-i\over k^2-M_W^2}+{-i\over k^2}{M_W^2\over M_W^2-k^2}.
\eqno(3.10)$$
As explained in Ref. \decomp, the first term gives rise to a massive
photon ($\gamma_{>}$) contribution.  It's effect combined with those from
$W$ and $Z$ will be refered to as the weak interaction contributions.
The second term is the same as for a massless photon $(\gamma_{<})$,
but with an additional cutoff $\Lambda=M_W$ as a regulator for
ultra-violet divergences.
The contribution of
$\gamma_{<}$ (Fig. 4a) and (Fig. 4b) plus the photonic box (Fig. 4c) (
evaluated with the full photon propagator $-i/k^2$) gives the
one-loop QED correction.  The result
for $\nu_e+e\to\nu_e+e$ is
$$\eqalign{\delta M^{QED}={-ig^2\over 2M_W^2}{\alpha\over 4\pi}&
\Bigl\{ [\bar{u}_e(p')\gamma_{\lambda}Lu_e(p')]
[\bar{u}_{\nu_e}(p)\gamma^{\lambda}Lu_{\nu_e}(p)]
{-q^2\over M_W^2}\Bigl({2\over 3}\ln {M_W^2\over m_e^2}-{13\over 9}\Bigr)
\cr
&+[\bar{u}_e(p')\gamma_{\lambda}\gamma_5u_e(p')]
  [\bar{u}_{\nu}(p)\gamma^{\lambda}Lu_{\nu}(p)]
\Bigl(1+{1\over 3}{q^2\over M_W^2}\Bigr)\Bigr\},\cr}\eqno(3.11)$$
where $q\equiv p-p'$.
We have omitted all terms directly proportional to $m_e$.

The result is finite in both the infrared and the ultra-violet region.
Only the axial current (the second term in Eq. (3.11))
 receives an order $\alpha$ correction.  This is because
QED does not introduce charge renormalization.  It is known that the
axial current term does not contribute to the neutrino coherent
forward scattering if the electrons in the medium are not
polarized. We will therefore neglect it from now on.

A notable feature of (3.11) is that it contains a mass
singular term $\ln M_W^2/m_e^2$. As we will see below,
this  mass singularity won't cancel even after
we add the weak interaction corrections.
This is not in conflict with the well known
Kinoshita-Sirlin-Lee-Naunberg (KSLN) theorem\rlap,\Ref\KSLN{
     T. Kinoshita and A. Sirlin, Phys. Rev. 113, 1652 (1959);
     T. Kinoshita, J. Math. Phys. 3, 650 (1962);
     T. D. Lee and M. Nauenberg, Phys. Rev. 133, 1549 (1962).}
however.
The KSLN theorem follows because of the finiteness of Green's
functions with non-exceptional Euclidean external momenta
in a renormalizable theory.  The non-exceptional momentum refers
to configurations in which no partial sum of momenta vanishes. Here,
we clearly have an exceptional external momentum configuration
if one allows $m_e\to 0$.  The mass singular term is multiplied by
$q^2$ which vanishes in the limit $m_{\nu_e}, m_e\to 0$ in the rest
frame of the electron.
 Therefore, the KSLN theorem does not apply
here.

Comparing with the tree level result (taking only the first term
of Eq. (2.6))
$$M_0\equiv {ig^2\over 2}{1\over M_W^2-q^2}
[\bar{u}_e(p')\gamma_{\alpha}Lu_{\nu_e}(p)]
[\bar{u}_{\nu_e}(p)\gamma^{\alpha}Lu_e(p')],\eqno(3.12)$$
the leading log
QED correction to the ratio $q^2/M_W^2$ that determines
the index of refraction (see Eq. (2.10))
is $-(\alpha/6\pi)\ln (M_W^2/m_e^2)$.
The dominant contribution is
provided by the wave function renormalization.
 Numerically,
it is $-0.9\%$.

\section{Weak Interaction Correction}

For reasons explained above, as far as the one-particle-reducible
graphs are concerned  only the $u$-channel scattering graphs
are of interest.  This makes our calculation very
similar to that of $\mu$-decay\rlap.\Ref\mudecay{A. Sirlin,
Phys. Rev. D22, 971 (1980).}  Accordingly, we will employ the
on-shell renormalization scheme of Sirlin (Ref. \mudecay)
$$\cos^2\theta_W={M_W^2\over M_Z^2}.\eqno(3.13)$$
For
convenience, we orgainize our results into
the self-energy, the vertex correction, and the box-diagram parts
$$\delta M^{Weak}=\delta M_{Self}+\delta M_{Vertex}+\delta M_{Box},
\eqno(3.14)$$
where $\delta M_{Self}$  and $\delta M_{Vertex}$ also
include the counterterm contributions and the wave function
renormalization effect, respectively.

The self-energy and counterterm diagrams are shown in Fig. 5.
The result for the sum of these graphs is known (Ref. \mudecay ):
$$\delta M_{Self}=M_0\Bigl[{A_{WW}(q^2)-ReA_{WW}(M_W^2)\over
q^2-M_W^2}-{2\delta e\over e}+{c^2\over s^2}Re\Bigl(
{A_{ZZ}(M_Z^2)\over M_Z^2}-{A_{WW}(M_W^2)\over M_W^2}\Bigr)\Bigr],
\eqno(3.15)$$
where $c\equiv \cos\theta_W$, $s\equiv \sin\theta_W$, and $M_0$
is given by Eq. (3.12).
The  functions $A_{WW}$ and $A_{ZZ}$ are the coefficients of
$g_{\mu\nu}$ in the $WW, ZZ$ self-energies.  These functions have been
calculated before.  We will use the results given by Marciano and
Sirlin\rlap.\Ref\selfenergy{W. J. Marciano and A. Sirlin,
Phys. Rev. D22, 2695 (1980).}

The vertex graphs are shown in Fig. 6.
In evaluating the wave function renormalization contributions we
use the massive photon propagator $\gamma_{>}$
 (the first term in Eq. (3.10)).
The result  is
$$\eqalign{\delta M_{6a+6b+6c+6d}=-M_0{\alpha\over 4\pi}
&\Bigl\{\Bigl({2\over \epsilon}+\Gamma'(1)-{1\over 2}
-\ln{M_Z^2\over 4\pi\mu_0^2}\Bigr)\Bigl({1\over 2c^2s^2}-1\Bigr)\cr
&+\Bigl({2\over\epsilon}+\Gamma'(1)-{1\over 2}
-\ln{M_W^2\over 4\pi\mu_0^2}\Bigr)\Bigl({1\over s^2}+1\Big)\Bigr\}.\cr}
\eqno(3.16)$$
To isolate the ultra-violet divergence, we have employed the
dimensional-regularization\Ref\dimension{
        G. 't Hooft and M. Veltman, Nucl. Phys. B44, 189 (1972).}
with $\mu_0^2$  the ultra-violent cut-off and $\Gamma(\epsilon /2)
=2/\epsilon+\Gamma'(1)+O(\epsilon)$.

Corrections to the vertex are obtained by evaluating the three-point
functions from
Fig. (6e) and Fig. (6f), in which one uses the full photon propagator.
We find (details can be found in the appendix)
$$\eqalign{\delta M_{6e+6f}=M_0{\alpha\over 2\pi}\Bigl\{&
\Bigl({2\over \epsilon}+\Gamma'(1)-\ln{M_W^2\over 4\pi\mu_0^2}\Bigr)
\Bigl({3\over s^2}+{s^2-c^2\over 4s^2c^2}\Bigr)\cr
&+{5\over 2}+{c^2\over s^2}\Bigl({5\over 2} +{3\over s^2}\ln c^2\Bigr)
-{s^2-c^2\over 4s^2c^2}\Bigl({1\over 2}-\ln c^2\Bigr)\cr
&-{q^2\over M_W^2}\Bigl[\ln{M_W^2\over m_e^2}+\Bigl(1-{1\over 2s^2}\Bigr)
F(q^2)+c'\Bigr]\Bigr\},\cr}\eqno(3.17)$$
where
$$c'={5\over 6}-{4c^2\over 3s^2}\Bigl(1-{1\over s^2}
+{2c^2\over s^4}\Bigr)
-{c^4\over s^4}\Bigl(1+{8c^2\over 3s^4}\Bigr)\ln c^2+{1\over 6s^2},
\eqno(3.18)$$
and
$$F(q^2)=\int^1_0dx 2x(x-1)\ln{xM_Z^2+(1-x)^2m_e^2\over
                               x(x-1)q^2+(1-x)m_e^2}.\eqno(3.19)$$
Combining Eqs. (3.16) and (3.17), we obtain
$$\eqalign{\delta M_{Vertex}=&
M_0{\alpha\over 4\pi}\Bigl\{{4\over s^2}\Bigl({2\over \epsilon}
+\Gamma'(1)-{1\over 2}-\ln {M_W^2\over 4\pi\mu_0^2}\Bigr)
+{c^2\over s^4}(5+c^2)\ln c^2+{8\over s^2}\cr
&-2{q^2\over M_W^2}
\Bigl[\ln{M_W^2\over m_e^2}+\Bigl(1-{1\over 2s^2}\Bigr)
F(q^2)+c'\Bigr]\Bigr\}.\cr}\eqno(3.20)$$
Comparing with the large log terms, one sees from Eq. (3.20)
that the constant $c'\approx 3$ clearly can be
ignored.  Since we are interested in regions in which $q^2$ is
space-like and
 $\vert q^2\vert \gg m_e^2$,
in the leading log approximation one can
make a further simplification
$$F(q^2)\approx {1\over 3}\ln {q^2\over M_Z^2}.\eqno(3.21)$$
Here, we have neglected the imaginary part that will be discussed
in detail in section 7.

There are  a total of eight box diagrams (Fig. 7).
A straightforward calculation
shows that the leading log  plus the order $\alpha$ terms  are given by
(details can be found in the appendix)
$$\eqalign{\delta M_{Box}=&{ig^2\over 2M_W^2}{\alpha\over 4\pi}
[\bar{u}_e(p')\gamma_{\alpha}Lu_e(p')]
[\bar{u}_{\nu_e}(p)\gamma^{\alpha}Lu_{\nu_e}(p)]\cr
&\times \Bigl[{5c^4-3s^4\over 2s^4}\ln c^2
             +{15-24s^2c^2\over 8s^2c^2}
             +{11+20s^2+56s^4\over 24s^2}
             {q^2\over M_W^2}\ln {q^2\over M_W^2}\Bigr]\cr
&+{ig^2\over 2M_W^2}{\alpha\over 4\pi}
[\bar{u}_e(p')\gamma_{\alpha}Lu_e(p')]
[\bar{u}_{\nu_e}(p)\gamma_{\beta}Lu_{\nu_e}(p)]
q^{\alpha}q^{\beta}{5-4s^2+8s^4\over 12s^2}\ln{q^2\over M_W^2}.\cr}
\eqno(3.22)$$
In reaching this simple result, we have made the approximation
$$I_1(m_{\nu_e},m_e, M_W; q^2)\approx -{1\over 6} {q^2\over M_W^2}
\ln{q^2\over M_W^2},\eqno(3.23)$$
and ignored all non-leading log terms.  The box diagrams also
generate terms which do not contribute to the neutrino
coherent forward scattering amplitudes.  They are of the form
$[\bar{u}_e\gamma_{\alpha}\gamma_5u_e][\bar{u}_{\nu_e}\gamma^{\alpha
}Lu_{\nu_e}]$ and $q^{\alpha}q^{\beta}
[\bar{u}_e\gamma_{\alpha}\gamma_5u_e]
[\bar{u}_{\nu_e}\gamma_{\beta}Lu_{\nu_e}] $ and have been
omitted from Eq. (3.22).  The result can be written in a more
compact form by using the relation
$$q^{\alpha}q^{\beta}[\bar{u}_e\gamma_{\alpha}u_e]
                     [\bar{u}_{\nu_e}\gamma_{\beta}Lu_{\nu_e}]
={q^2\over 2}[\bar{u}_e\gamma_{\alpha}u_e]
             [\bar{u}_{\nu_e}\gamma^{\alpha}Lu_{\nu_e}].\eqno(3.24)$$
It then follows that
$$\eqalign{\delta M_{Box}=&{ig^2\over 2M_W^2}{\alpha\over 4\pi}
[\bar{u}_e(p')\gamma_{\alpha}Lu_e(p')]
[\bar{u}_{\nu_e}(p)\gamma^{\alpha}Lu_{\nu_e}(p)]\cr
&\times \Bigl({5c^2-3s^4\over 2s^4}\ln c^2+{15-24s^2c^2\over 8s^2c^2}
+{2\over 3s^2}(1+s^2+4s^4){q^2\over M_W^2}\ln {q^2\over M_W^2}\Bigr)\cr
&+...,\cr}
\eqno(3.25)$$
where the ellipses represent those (incoherent) terms of no interest to
us.

Among the eight box diagrams,
the first four also appear in $\mu$-decay (with appropriate
change of external lines).  The order $\alpha$ correction
of these graphs is given by the first term of Eq. (3.25).
{}From Eqs. (3.20) and (3.25), one can see that the coefficients of
$q^2/M_W^2$ in $\delta M_{Vertex}$ and in $\delta M_{Box}$
are almost equal  but with the opposite  sign.  This reults in a
large cancellation between these two contributions.

\section{Combination of Results}

The combination of all weak interaction corrections plus the
tree level result (Eq. (3.12)) can be expressed in terms
of the weak coupling constant
 $G_{\mu}=(1.16639\pm 0.00002)\times 10^{-5}\ GeV^{-2}$
determined very accurately from $\mu$-decay by the
substitution (Ref.\mudecay )
$${g^2\over M_W^2}(1+\Delta r)={8G_{\mu}\over \sqrt 2},\eqno(3.26)$$
where
$$\eqalign{\Delta r=&{ReA_{WW}(M_W^2)-A_{WW}(0)\over M_W^2}
-{2\delta e\over e}+{c^2\over s^2}\Bigl({A_{ZZ}(M_Z^2)\over M_Z^2}
-{A_{WW}(M_W^2)\over M_W^2}\Bigr)\cr
&+{\alpha\over 4\pi}\Bigl[ {4\over s^2}\Bigl({2\over\epsilon}+\Gamma'(1)
-\ln{M_Z^2\over 4\pi\mu_0^2}\Bigr)+\Bigl({7\over 2}-6s^2\Bigr)
{\ln c^2\over s^4}+{6\over s^2}\Bigr].\cr}\eqno(3.27)$$
It then follows from Eqs. (3.12), (3.14), (3.20), (3.25) and (3.26) that
$$\eqalign{M_0+\delta M^{Weak}=-i2\sqrt 2&G_{\mu}
[\bar{u}_e(p')\gamma_{\alpha}Lu_e(p')]
[\bar{u}_{\nu_e}(p)\gamma^{\alpha}Lu_{\nu_e}(p)]
\Bigl\{1-{\alpha\over 4\pi}\Bigl({15-24s^2c^2\over 8s^2c^2}\Bigr)\cr
&+{q^2\over M_W^2}\Bigl[1+\Delta R(q^2)+{\alpha\over 4\pi}\Bigl(
{1+4s^2+8s^4\over 3s^2}\ln {M_W^2\over q^2}-2\ln {M_W^2\over m_e^2}
\Bigr)\Bigr]\Bigr\},\cr}\eqno(3.28)$$
where
$$\eqalign{\Delta R(q^2)\equiv &{M_W^2\over q^2}
\Bigl({ReA_{WW}(M_W^2)-A_{WW}(q^2)\over M_W^2-q^2}
-{ReA_{WW}(M_W^2)-A_{WW}(0)\over M_W^2}\Bigr)\cr
&={A_{WW}(0)-A_{WW}(q^2)\over q^2}+{ReA_{WW}(M_W^2)-A_{WW}(0)\over
M_W^2}+O(M_W^{-4})\cr}\eqno(3.29)$$
is the $W$ self-energy contribution, with the understanding that only
the real part of the result will be kept.

The residual order $\alpha$ correction
 (the second term in Eq. (3.28)) arises because we have omitted
all the $t$-channel one-particle-reducible graphs. Also, the last
four box diagrams do not appear in the calculation of
$\mu$-decay.  In any case,
this term does not contribute to the index of refraction.

Terms in the squared brackets will contribute to
the neutrino index of refraction.  The first term is the tree level
result obtained by the expansion of the $W$ propagator.  The second
term, $\Delta R(q^2)$,
is due to  $W$ self-energy and counterterms, and the rest is
the combination of vertex and box diagram corrections.

The large cancellation between the vertex  and the
box diagrams  makes their contribution rather small.
The sum is about $+2\%$ for $\vert q^2\vert=1\ MeV^2$ and it becomes
negligibly small, $+0.1\%$, for $\vert q^2\vert=400\ MeV^2$.

The largest contribution comes
from $W$ self-energy.  For convenience, we will decompose it into
the bosonic, the hadronic and the leptonic parts
$$\Delta R(q^2)=\Delta R^{(b)}(q^2)+\Delta R^{(h)}(q^2)
+\Delta R^{(\ell)}(q^2).\eqno(3.30)$$
The bosonic contribution is  negligible, because
it does not have any large log term.  Explicitly,
from Eq. (3.29) and the given self-energy function in Ref. \selfenergy\
we find
$$\Delta R^{(b)}(q^2)
={\alpha\over 4\pi}\Bigl({65\over 18}-{3c^2\over 2s^2}
+{1\over s^2}\int^1_0dxF(x,\xi)\Bigr),\eqno(3.31)$$
where
$$\eqalign{F(x,\xi)=&\Bigl[{2s^4-16c^2-10c^4-1\over 2c^2}
+{1+4c^4+16c^2\over 2c^2}x -{20c^2+1\over 2}x^2\Bigr]\ln
{c^2x^2-x+1\over c^2x-x+1}\cr
&+{2-\xi+\xi x-x^2\over 2}\ln{x^2+\xi (1-x)\over x+\xi (1-x)}
+{x(1-x)\over x+\xi (1-x)}+{s^4x(1-x)\over c^2x^2+(1-x)},\cr}\eqno(3.32)
$$
and $\xi\equiv m_{\phi}^2/M_W^2$.   Here $m_{\phi}$ is the Higgs mass.

The hadronic correction seems to provide
the biggest contribution, but its calculation is complicated by
strong interaction physics.  The conventional wisdom here is to
employ dispersion relations to relate $\Delta^{(h)}(q^2)$ to some
physically measurable quantities.  While a detailed study of this
problem lies beyond the scope of the present paper, we will estimate
the result by employing  the constituent quark mass.
We find for $\vert q^2\vert
\ll m_i^2$ (here $m_i$ is a constituent quark mass)
$$\eqalign{
\Delta R^{(h)}(q^2)={3\alpha\over 2\pi s^2}&\sum_{i,j}\vert V_{ij}\vert^2
Re\int^1_0dx\Bigl[x(1-x)\ln{m_i^2x+m_j^2(1-x)-M_W^2x(1-x)\over
            m_i^2x+m_j^2(1-x)}\cr
&-{m_i^2x+m_j^2(1-x)\over 2M_W^2}\ln{m_i^2x+m_j^2(1-x)-M_W^2x(1-x)\over
m_i^2x+m_j^2(1-x)}-{1\over 12}\Bigr],\cr}\eqno(3.33)$$
where $V_{ij}$ is the KM matrix element\rlap.\Ref\KM{
      M. Kobayashi and T. Maskawa, Prog. Theor. Phys. 49, 652 (1973).}
The sum is over all three families of quarks with
$m_{u,d}\approx 300\ MeV$, $m_s\approx 450\ MeV$, etc.
The second term in Eq. (3.33) is obviously negligible for the first two
generations.  We present it here is to show that it is also
negligible for a heavy quark.  Indeed, for $m_t\gg M_W$, it becomes
$x(1-x)/2$, which is much smaller than the first leading log term.
Numerically, we find that  $\Delta R^{(h)}(q^2)$ is about $+5\%$.

The calculation of the leptonic contribution
is much less ambiguous.  We find that the leading log contribution is
(neglecting possible mixings in the lepton sector)
$$\Delta R^{(\ell)}(q^2)
={\alpha\over 2\pi s^2}\sum_{\ell=e,\mu,\tau}
Re\int^1_0dxx(1-x)\ln{m_{\ell}^2-M_W^2(1-x)\over m_{\ell}^2-q^2(1-x)}.
\eqno(3.34)$$
Numerically, it varies from $+4\%$ to $+3\%$ in the region
$1\ MeV^2\le \vert q^2\vert\le 400\ MeV^2$.

Now, the sum of all weak corrections plus a $-1\%$ QED
correction varies from about $+10\%$ to about $+7\%$ in the region
of interest discussed above.  Our results are summarized in Table 1.

The result for $\nu_e+\bar{e}\to \nu_e+\bar{e}$ can be obtained
from that of neutrino-electron scattering by changing
$\bar{u}_e\gamma_{\alpha}Lu_e$ to $-\bar{u}_e\gamma_{\alpha}Ru_e$
and $q^2\to -q^2$ (valid only if the neutrino and the lepton mass is
negligible).
The outcome is that the constant term ($q^2$
independent) has the opposite sign, whereas the $q^2/M_W^2$ term
remains the same.

To obtain the index of refraction, we now need to take the thermal
average of Eq. (3.28).  Since the variation of radiative corrections
in the region of $q^2$ of interest is not significant, in the leading
log approximation we can write the index of refraction in a
$N_e=N_{\bar{e}}$ medium as
$$(n-1)_{\nu_ee(\bar{e})\to \nu_ee(\bar{e})}=-{16\sqrt 2\over
3M_W^2}G_{\mu}N_ep_0\langle p_0'\rangle
(1+\delta_{\nu_e e}),\eqno(3.35)$$
where $\delta_{\nu_e e}
$ is simply given by the non-averaged result discussed
above; it
 is between about $+10\%$ to about $+7\%$ (see Table 2 for detail).
Results for other values of $q^2$ can easily be obtained
from the analytic formulae given above and in the appendix,
provided the leading log approximation can still be justified.

\bigskip
\chapter{Neutrino-(anti)neutrino Scattering}

Fig. 8 shows the tree level Feynman graphs for a diagonal (in family
space)
neutrino-neutrino and a neutrino-anti-neutrino scattering.
Fig. 8a and 8c are the standard $t$-channel
forward scattering graphs
induced by the neutral current interaction, where the coherent
condition requires that the virtual $Z$ cannot  carry any momentum.
This feature is shared by all  $t$-channel one-particle-reducible
graphs.  As explained before, the sum of coherent amplitudes of
these $t$-channel graphs
cancels for $\nu+\nu\to \nu+\nu$ plus
$\nu+{\bar{\nu}}\to\nu+{\bar{\nu}}$.  Therefore, we will ignore them
from now on.

Due to the identical particle property, there is another angle for which
the  $\nu\nu\to\nu\nu$ scattering can  be coherent.
This corresponds to the configuration in which the projectile and
the scatterer switch places after  scattering (Figs. (8b)).
In contrast to the usual
$t$-channel scattering, now the coherent condition
allows the virtual $Z$ to have an arbitrary momentum $ q^2\equiv
(p- p'^2)$

Analogous to the Bhabha scattering in QED,  there is an
$s$-channel graph for the
$\nu{\bar{\nu}}\to\nu{\bar{\nu}}$ scattering (Fig. 8d), in which
the virtual $Z$ carries a momentum  $(p+p')^2=-q^2$.
Again, the momenta of the virtual $Z$s in Fig. 8b and Fig. 8d have
the opposite sign, and thus
their sum  survives the cancellation that
occurs in lowest order.

\section{$\nu_e+\nu_e(\bar{\nu}_e)\to \nu_e+\nu_e(\bar{\nu}_e)$}

For definiteness, we now consider
the $\nu_e\nu_e\to\nu_e\nu_e$ scattering.  Of cosmological interest,
we consider situations  where
$\vert q\vert$ is of the order of a few $MeV$.  The tree level
matrix element of Fig. 8b is simply
$$M_0=-\Bigl({ig\over 2c}\Big)^2{i\over M_Z^2-q^2}
[\bar{u}_{\nu_e}(p')\gamma_{\alpha}Lu_{\nu_e}(p)]
[\bar{u}_{\nu_e}(p)\gamma^{\alpha}Lu_{\nu_e}(p')],\eqno(4.1)$$
where $p'$ and $p$ refer to the momenta of the scatterer and
the projectile, respectively.  The over all minus-sign is due to
the exchange of identical fermionic particles.
Following the discussion of
section 2, one finds that the  index of refraction arising
from  neutrino-neutrino and neutrino-anti-neutrino scattering
in a medium with equal amount of neutrinos and anti-neutrinos
is
$$(n-1)_{\nu_e+\nu_e(\bar{\nu}_e)\to\nu_e+\nu_e(\bar{\nu}_e)}
={\cos^2\theta_W
\over 4}(n-1)_{\nu_ee(\bar{e})\to\nu_ee(\bar{e})},\eqno(4.2)$$
where $(n-1)_{\nu_e+e
(\bar{e})\to\nu_e+e(\bar{e})}$ is given by Eq. (2.10).

Because of the absence of QED corrections, the calculation of one-loop
radiative corrections  to  neutrino-neutrino scattering is
considerably simpler.  However,
 in the case of $\nu_ee\to\nu_ee$
scattering we are interested in  radiative corrections to the charged
current, whereas here the interest is shifted to the neutral current
interaction.  The result will therefore have a
moderate dependence on the top quark mass.
For convenience, we again orgainize our results
into the self-energy, the vertex correction, and the box-diagram parts
$$\delta M^{Weak}=\delta M_{Self}+\delta M_{Vertex}+\delta M_{Box}.
\eqno(4.3)$$

The result for the self-energy and counterterm (Fig. 9) can  be obtained
from the analysis of neutral current radiative corrections of
Marciano and Sirlin (Ref. \selfenergy )
$$\delta M_{Self}=M_0\Bigl[{A_{ZZ}(q^2)-ReA_{ZZ}(M_Z^2)\over
q^2-M_Z^2}-{2\delta e\over e}+\Bigl({c^2\over s^2}-1\Bigr)
Re\Bigl({A_{ZZ}(M_Z^2)\over M_Z^2}-{A_{WW}(M_W^2)\over M_W^2}\Bigr)\Bigr]
,\eqno(4.4)$$
where $M_0$ is given by Eq. (4.1).

Neglecting the neutrino and the electron mass, we find that
the leading log vertex correction (Fig. 10) is
$$\delta M_{Vertex}=M_0{\alpha\over 4\pi}\Bigl[ 4{c^2\over s^2}\Bigl(
{2\over\epsilon}+\Gamma'(1)-\ln{M_W^2\over 4\pi\mu_0^2}\Bigr)+
{q^2\over M_W^2}\Bigl({4s^2-1\over 3s^2}\Bigr)\ln{M_W^2\over q^2}\Bigr].
\eqno(4.5)$$
Here, we have made the approximation indicated by Eq. (3.23) which is
valid for our purpose.
A more complete result
without this constraint can be found in the appendix.  In any case,
one can see from Eq. (4.5) that the vertex correction
to the neutrino index of refraction
 is completely
negligible
because $4s^2-1$ is close to zero.

Finally, there are a total of six box diagrams (Fig. 11).
The sum of their leading log contributions is
$$\delta M_{Box}=M_0{\alpha\over 4\pi s^2}\Bigl(2-{3\over 2c^2}
+{11\over 6}{q^2\over M_W^2}\ln {M_W^2\over q^2}\Bigr).\eqno(4.6)$$
In reaching this result we have again made use of the approximation
of Eq. (3.23).  Also, non-coherent terms similar to those
encountered in the above section have been excluded.
It then follows that in the renormalization scheme discussed above
$$\eqalign{M_0+\delta M^{Weak}=-i\sqrt 2G_{\mu}\rho^{(\nu;\nu)}&
[\bar{u}_{\nu_e}(p)\gamma_{\alpha}Lu_{\nu_e}(p)]
[\bar{u}_{\nu_e}(p')\gamma^{\alpha}Lu_{\nu_e}(p')]\cr
&\times\Bigl\{1+{q^2\over M_Z^2}\Bigl[1+\Delta R'(q^2)+
{\alpha\over 4\pi c^2s^2}{9+8s^2\over 6}
\ln{M_W^2\over q^2}\Bigr]\Bigr\},\cr}\eqno(4.7)$$
where  $\Delta R'(q^2)$ is the analogue of $\Delta R(q^2)$ in
neutrino-electron scattering
$$\Delta R'(q^2)={A_{ZZ}(0)-A_{ZZ}(q^2)\over q^2}+{A_{ZZ}(M_Z^2)-
A_{ZZ}(0)\over M_Z^2}+O(M_W^{-4}).\eqno(4.8)$$
Again  only the real part will be kept.  The over all
normalization constant $\rho^{(\nu;\nu)}$ can be conveniently
related to the constant $\rho_{N.C.}^{(\nu;\ell)}$
introduced in the study of $\nu\ell\to\nu\ell$ scattering
(Refs. \selfenergy , \SSM )
$$\rho^{(\nu;\nu)}=\rho^{(\nu;\ell)}_{N.C.}-{\alpha\over 4\pi}
\Bigl({1\over 2s^2c^2}\Bigr)\Bigl({7\over 2}
-10s^2+12s^4\Bigr),\eqno(4.9)
$$
where
$$
\eqalign{&\rho^{(\nu;\ell)}_{N.C.}=1+{\alpha\over 4\pi}\Bigl(
{3\over 4s^4}\ln c^2-{7\over 4s^2}+{2C_Z\over c^2s^2}+G(\xi,c^2)
+{3m_t^2\over 4s^2M_W^2}\Bigr),\cr
&G(\xi,c^2)={3\xi\over 4s^2}\Bigl({\ln (c^2/\xi)\over c^2-\xi}
+{1\over c^2}{\ln\xi\over 1-\xi}\Bigr),\cr
&C_Z={19\over 8}-{7\over 2}s^2+3s^4,\cr}\eqno(4.10)$$
and  $\xi=m_{\phi}^2/M_Z^2$.

 Again, the order $\alpha$ correction (the first term in Eq. (4.7))
cancels for the sum of neutrino-neutrino and neutrino-anti-neutrino
scattering.  Terms surviving the cancellation are grouped in the squared
brackets.  Among them,
 the first term is due to the expansion of the $Z$
propagator, the second is due to the $Z$ self-energy and the last term
is the sum of vertex and box diagram corrections.
 The
difference between
$\rho^{(\nu;\nu)}$ and $\rho^{(\nu;\ell)}_{N.C.}$ arises
because the three exchange box-diagrams (Fig. 11d to Fig. 11f)
do not appear in a $\nu\ell\to \nu\ell$ scattering.  Numerically, this
difference is very small, and
can be ignored
in our leading log approximation.
The aforementioned top quark mass dependence is included in  the
parameter
$\rho^{(\nu;\ell)}_{N.C.}$ (Eq. (4.10)).  This gives an
additional correction of the order
$+1\%$ for a heavy top with mass
$150\ GeV\lsim m_t\lsim 200\ GeV$.

In contrast to the $\nu_ee\to\nu_ee$ scattering,
no significant cancellation between the vertex correction and the
box diagram correction
 occurs.    As we
already pointed out, this is because the vertex correction is
suppressed by  $4s^2-1$.  As a result,
within the region $1\ MeV^2\le \vert q^2\vert \le
400\ MeV^2$,  $\delta M_{Vertex}+\delta M_{Box}$
is essentially given by $\delta M_{Box}$. Numerically, we find that
it is
about $+13\%$  to  $+10\%$.

For the self-energy contributions, we again decompose them
into the bosonic,  hadronic, and  leptonic parts
$$\Delta R'(q^2)=\Delta R^{\prime (b)}(Q^2)
+\Delta R^{\prime (h)}(q^2)+\Delta R^{\prime (\ell)}(q^2).\eqno(4.11)$$
The bosonic part does not have any large log terms, and hence
its contribution is negligible.  Explicitly,
$$\eqalign{\Delta R^{\prime (b)}(q^2)=-{\alpha\over 4\pi s^2}&
\Bigl[{1\over 6}\Bigl({17c^2\over 2}+{1+s^4\over 2c^2}-s^2+2s^4c^2\Bigr)
+{1\over c^2}\int^1_0dx{x(x-1)\over x(1-\xi)+\xi}\cr
&+\Bigl({17\over 2}+{s^4\over 2c^4}-{s^2\over c^2}-2s^4+5c^2\Bigr)
\int^1_0dx\ln {c^2+x(x-1)\over c^2}\cr
&+\Bigl({23c^2\over 2}+{s^4\over 2c^2}-1\Bigl)\int^1_0
dxx(x-1)\ln{c^2+x(x-1)\over c^2}\cr
&+{1\over 2c^2}\int^1_0dx\Bigl(x^2+(1-x)\xi-2\Bigr)\ln
{x^2+\xi (1-x)\over x+\xi (1-x)}\Bigr].\cr}\eqno(4.12)$$
Numerically, we find that the terms in the square
brackets add up to a value of  the order of $-1$ and they are
 not sensitive
to the choice of $\xi$.
Thus,
$\Delta R^{\prime (b)}(q^2)\lsim 0.3\%$.

The hadronic part  suffers the same ambiguity as $\Delta R^{(h)}(q^2)$
induced by the complication of strong interaction.  Within
the constituent quark approximation, we find
$$\Delta R^{\prime (h)}(q^2)=
{3\alpha\over 2\pi c^2s^2}
\sum_{f}\Bigl({1\over 2}-2s^2C_{3f}Q_f+4s^4Q_f^2\Bigr)
Re\int^1_0dxx(1-x)\ln{m_f^2-M_Z^2x(1-x)\over m_f^2-q^2x(1-x)},
\eqno(4.13)$$
where $Q_f$ and $C_{3f}$ are the charge and the third component
of weak isospin of the quark $f$ with a constituent mass $m_f$.
Again, it is easy to show that the contribution due to a heavy quark
is completely negligible in $\Delta R^{\prime (h)}(q^2)$.
Numerically, one finds from Eq. (4.13) that for
$\vert q^2\vert \ll m_t^2$,
$\Delta R^{\prime (h)}(q^2)$  is  about
$+5\%$.

The leptonic contribution, $\Delta R^{\prime (\ell)}$ can be
obtained from Eq. (4.13) by  an appropriate change of
charge and weak isospin quantum numbers for the charged leptons
and neutrinos, and then dividing the result by  3.
The sum is over all the three families of leptons and neutrinos.
  In the region
$1\ MeV^2\le \vert q^2\vert \le 400\ MeV^2$, we find that
$\Delta R^{\prime (\ell)}(q^2)$ is between $+7\%$ and $+5\%$.

Thus, the sum of all weak corrections plus a $+1\%$ correction if
the top quark mass is within the region of  $150\ GeV$ to $200\ GeV$ is
quite sizable $(\sim +20\%)$.  The details are summarized in Table 1,
in which one finds that for $1\ MeV^2 \le \vert q^2\vert \le 400 MeV^2$
the total correction is between $+25\%$ to $+20\%$.

The result for $\nu_e\bar{\nu}_e\to\nu_e\bar{\nu}_e$ can
be obtained from that of
 $\nu_e(p){\nu}_e(p')\to\nu_e(p){\nu}_e(p')$ by a
change of $\bar{u}(p')\gamma_{\alpha}Lu(p')\to-\bar{u}(p')\gamma_{\alpha}
Ru(p')$ and replacing $q^2$ by $-q^2$.
While the order $\alpha$ constant term will cancel
for the sum of these two amplitudes,  terms proportional to
$q^2/M_Z^2$ add up.  Thus, the index of refraction due to
$\nu_e\nu_e(\bar{\nu}_e)\to\nu_e\nu_e(\bar{\nu}_e)$ scattering
in a medium which contains an equal amount of $\nu_e$ and $\bar{\nu}_e$
is
$$(n-1)_{\nu_e\nu_e(\bar{\nu}_e)\to\nu_e\nu_e(\bar{\nu}_e)}
=-{8\sqrt 2\over 3M_Z^2}G_{\mu}
\rho^{(\nu;\nu)}N_{\nu_e}p_0\langle p_0'\rangle (1+\delta_{\nu_e\nu_e})
.\eqno(4.14)$$
In the leading log approximation, $\delta_{\nu_e\nu_e} $ is about $+20\%$
to $+25\%$
in the region $1\ MeV^2\le \vert q^2\vert \le 400\ MeV^2$.
The details are summarized in Table 2.

\section{$\nu_{\mu,\tau}+\nu_{\mu,\tau}(\bar{\nu}_{\mu,\tau})
\to \nu_{\mu,\tau}+\nu_{\mu,\tau}(\bar{\nu}_{\mu,\tau})$}

In the same region of $q^2$ discussed above, we now have a slightly different
hierarchy
$$M_{W,Z}^2\gg m_{\mu,\tau}^2\gg \vert q^2\vert.\eqno(4.15a)$$
However, the  log factor $\ln (M_W^2/m_{\ell}^2)\ (\ell=\mu,\tau)$
is still rather large, and thus a leading log approximation calculation
is still meaningful.  In this case the function $I_1$ defined in
Eq. (3.1) can be approximately simplified to
$$I_1(m_{\nu_{\ell}},m_{\ell}, M_W, q^2)
\approx -{1\over 6}\ln{m_{\ell}^2\over M_W^2}.\eqno(4.15b)$$
As a result, some of the log factors in $\delta M_{Vertex}$
and $\delta M_{Box}$ in Eqs. (4.5) and (4.6) should be changed
from $\ln (q^2/M_W^2)$ to $\ln (m_{\ell}^2/M_W^2)$.  The rest of
the calculation remains the same.  Places where such a change should
be made are those in which  diagrams  contain charged leptons.

For the vertex correction we find
$$\delta M_{Vertex}=M_0{\alpha\over 4\pi}\Bigl[ {4c^2\over s^2}
\Bigl({2\over\epsilon}+\Gamma'(1)-\ln{M_W^2\over 4\pi\mu_0^2}\Bigr)
+{q^2\over M_W^2}\Bigl({4s^2-2\over 3s^2}\ln{M_W^2\over m_{\ell}^2}
+{1\over 3s^2}\ln{M_Z^2\over q^2}\Bigr)\Bigr],\eqno(4.16)$$
where
$$M_0={ig^2\over 4c^2}{1\over M_Z^2-q^2}
[\bar{u}_{\nu_{\ell}}(p')\gamma_{\alpha}Lu_{\nu_{\ell}}(p)]
[\bar{u}_{\nu_{\ell}}(p)\gamma^{\alpha}Lu_{\nu_{\ell}}(p')].\eqno(4.17)$$
Eq. (4.16) reduces to  Eq. (4.5) if one substitues
$m_{\ell}^2$ by $q^2$ and ignores the difference (valid in the leading
log approximation) between $M_W$ and $M_Z$.

Accordingly, the box diagram contribution becomes
$$\delta M_{Box}=M_0{\alpha\over 4\pi s^2}\Bigl[2-{3\over 2c^2}
+{q^2\over M_W^2}\Bigl({2\over 3}\ln{M_W^2\over m_{\ell}^2}
+{7\over 6}\ln{M_W^2\over q^2}\Bigr)\Bigr].\eqno(4.18)$$

It then follows from Eqs. (4.16) and (4.18) that for
$\nu_{\ell}+\nu_{\ell}\to\nu_{\ell}+\nu_{\ell}\ (\ell=\mu,\tau)$
our leading
log result is
$$\eqalign{M_0+\delta M^{Weak}=&-i\sqrt 2G_{\mu}\rho^{(\nu;\nu)}
[\bar{u}_{\nu_{\ell}}(p)\gamma_{\alpha}Lu_{\nu_{\ell}}(p)]
[\bar{u}_{\nu_{\ell}}(p')\gamma^{\alpha}Lu_{\nu_{\ell}}(p')]\cr
&\times\Bigl\{1+{q^2\over M_Z^2}\Bigl[1+\Delta R^{\prime}(q^2)
+{\alpha\over 4\pi s^2c^2}\Bigl({9+8s^2\over 6}\ln{M_W^2\over q^2}
+{8s^2\over 6}\ln{q^2\over m_{\ell}^2}
\Bigr)\Bigr]\Bigr\}.\cr}\eqno(4.19)$$
It bares a strong resemblance to Eq. (4.7) for
$\nu_e+\nu_e\to \nu_e+\nu_e$ scattering.

Given the smallness of $s^2$, the difference between Eqs. (4.7) and
(4.19) is almost purely academic.  Thus, qualitatively, we expect
that the total correction for the three types of neutrino-neutrino
scattering are approximately the same ($\sim +20\%$).
The details are summarized in Table 2.

\bigskip

\chapter{Scattering Between Different Families}

In the absence of family mixing,   lowest order interactions
between different families such as
$\nu_{\mu,\tau}+e(\bar{e})\to\nu_{\mu,\tau}+e(\bar{e})$ and
$\nu_i+\nu_j(\bar{\nu}_j)\to \nu_i+\nu_j(\bar{\nu}_j) (i\ne j)$
 can only
occur through the $t$-channel.
Then, for  the reasons  explained above, the coherent amplitudes of
these interactions will cancel completely, and  they will
not contribute to the neutrino index of refraction.
The situation will change if one goes
beyond the lowest order.  In this case, radiative corrections
become the only contribution.

\section{$\nu_{\mu,\tau}+e(\bar{e})\to\nu_{\mu,\tau}+e(\bar{e})$}

It should be clear by now that in the absence of family mixing only
those diagrams which are similar to
the last four  in Fig. 7  contribute to the neutrino
index of refraction.  The only difference is to change the
$\nu_e$ line by a $\nu_{\mu,\tau}$ line, and, accordingly,
the internal electron line by a $\mu$ or $\tau$ line.
The result  is
$$M_{\nu_{\mu,\tau}+e\to\nu_{\mu,\tau}+e}=M_0
{\alpha\over 4\pi s^2}\Bigl\{ {15-24s^2c^2\over 4c^2}+
{q^2\over M_W^2}\Bigl[ {10-8s^2+16s^4\over 3}\ln {q^2\over M_W^2}
+{8\over 3}\ln{m_{\mu,\tau}^2\over q^2}\Bigr]\Bigr\}+...,\eqno(5.1)$$
where the ellipses represent  incoherent terms which are of
no interest to us, and
$$M_0={ig^2\over 4M_W^2}[\bar{u}_e(p')\gamma_{\alpha}Lu_e(p')]
[\bar{u}_{\nu_{\mu,\tau}}(p)\gamma^{\alpha}Lu_{\nu_{\mu,\tau}}(p)].
\eqno(5.2)$$
In Eq. (5.1) terms proportional to $q^2/M_W^2$ contribute to  the
neutrino index of refractions.  By employing the by-now familiar method,
we find
$$(n-1)_{\nu_{\ell}+e(\bar{e})\to\nu_{\ell}+e(\bar{e})}
=-{8\sqrt 2\over 3M_W^2}G_{\mu}N_ep_0
\langle p_0'\rangle\delta_{\nu_{\ell}e},\ \ \ \
\ell=\mu,\tau.\eqno(5.3)$$
In the region $1\ MeV^2 \le \vert q^2\vert \le 400\ MeV^2$,
$\delta_{\nu_{\mu}e}$ varies from about $+11\%$ to $+10\%$, and
$\delta_{\nu\tau}$  is between $+7\%$ to $+6\%$ (see Table 2).
Notice also,
$N_e=2N_{\nu}$ (see Eq. (1.1)).  Such a sizable correction in the
diagonal scattering $(\nu_ee\to\nu_ee)$ is cancelled by the vertex
correction.  There is no vertex correction for scatterings
between different families if the corresponding mixing is zero.

\section{$\nu_i+\nu_j(\bar{\nu}_j)
\to\nu_i+\nu_j(\bar{\nu}_j) (i\ne j)$}

Leading log contributions are generated from box diagrams
similar to the first three graphs in Fig. 11.  A straightforward
calculation shows that in the zero mixing limit
$$\eqalign{M=&{-ig^2\over 4M_W^2}
[\bar{u}_{\nu_i}(p)\gamma_{\alpha}Lu_{\nu_i}(p)]
[\bar{u}_{\nu_j}(p')\gamma^{\alpha}Lu_{\nu_j}(p')]\cr
&\times{\alpha\over 4\pi s^2c^2}\Bigl[c^2-{3\over4}+
{q^2\over M_Z^2}\Bigl({5\over 12}\ln{M_Z^2\over m_{\ell_j}^2}
+{2\over 3}\ln{M_Z^2\over q^2}\Bigr)\Bigr]+...,\cr}\eqno(5.4)$$
where the ellipses again refer to the incoherent terms and
$m_{\ell_j}$ is the heaviest charged lepton mass in question.
It then follows that
$$(n-1)_{\nu_i+\nu_j(\bar{\nu}_j)\to\nu_i+\nu_j(\bar{\nu}_j)}
=-{8\sqrt 2\over 3M_W^2}G_{\mu}N_{\nu_j}
p_0\langle p_0'\rangle\delta_{\nu_i\nu_j}.\eqno(5.5)$$
Numerically, we find that in the region $1\ MeV^2\le \vert q^2\vert
\le 400\ MeV^2$, $\delta_{\nu_i\nu_j}$ is about $+6\%$ to $+5\%$ for
the scattering $\nu_{e,\mu}+\nu_{\tau}\to\nu_{e,\mu}+\nu_{\tau}$.
The result increases about $+1\%$ for
$\nu_e+\nu_{\mu}\to\nu_e+\nu_{\mu}$.  The details are summarized in Table
2.

\bigskip
\chapter{Neutrino-photon Scattering}

Since a photon is its own anti-particle, the cancellation
discussed above does not apply to the neutrino-photon scattering.
However, we will show below that the neutrino-photon
forward scattering amplitude is zero at one-loop level
in the standard model.

It is known\Ref\twophoton{Jiang Liu, Phys. Rev. D44, 2879, (1991),
    and references there in.} that to the lowest nonvanishing
order the diagonal effective neutrino-two photon interaction
is given by
$${\it{L}}_{eff}=a\bar{\nu}\nu F^{\alpha\beta}F_{\alpha\beta}
+ia'\bar{\nu}\gamma_5\nu \tilde{F}^{\alpha\beta}F_{\alpha\beta},
\eqno(6.1)$$
where $\nu$ is a neutrino field, $F^{\alpha\beta}$ is the
electromagnetic tensor with its dual $\tilde{F}^{\alpha\beta}=
{1\over 2}\epsilon^{\alpha\beta\sigma\lambda}F_{\sigma\lambda}$.
In terms of the standard electroweak interactions, $a$ and $a'$
are zero at one-loop level if neutrinos are massless.  Assuming
neutrinos are massive, one then finds (Ref. \twophoton )
$$a'={G_F\alpha\over \sqrt \pi}m_{\nu} f,\ \ \ \  a=0,\eqno(6.2)$$
where $m_{\nu}$ is a neutrino mass and $f$ is a scalar function
obtained from  loop integrals with its
real part   given by
$$Ref={-1\over 4k_1.k_2}+{m_{\ell}^2\over 4(k_1.k_2)^2}\times
\cases{2\Bigl[\sin^{-1}\sqrt{k_1.k_2/2m_{\ell}^2}\Bigr]^2,
&if $k_1.k_2<2 m_{\ell}^2$\cr
{\pi^2\over 2}-{1\over 2}\ln^2{1+\sqrt{1-2m_{\ell}^2/k_1.k_2}\over
                              1-\sqrt{1-2m_{\ell}^2/k_1.k_2}}
,&if $k_1.k_2\ge 2m_{\ell}^2$.\cr}\eqno(6.3)$$
Here, $k_1$ and $k_2$ are, respectively,
 the initial- and the final-photon
momentum.

The function $f$ has a notable feature that it is zero
for a forward scattering in which  $k_1.k_2=0$.  Thus,
in the forward direction  ${\it{L}}_{eff}$ is zero
at one-loop level.  Notice also, the matrix element
$\bar{u}(p)_{\nu}\gamma_5u(p)_{\nu}$ is zero.  We then conclude
that the
 neutrino-photon scattering does not contribute to
the real part of the neutrino index of refraction at one-loop level.
Thus, the photon plasma is essentially transparent to   neutrinos.

\bigskip

\chapter{The Imaginary Part of the Index of Refraction}

While the real part of the index of refraction describes
the coherent interference of propagating neutrinos, the
imaginary part characterizes the incoherent
depletion of neutrinos from their original
coherent state.
In normal situations, $Re(n)\gg Im(n)$, because neutrinos only
participate in weak interactions and hence the phase shift in
the forward scattering amplitude is very small.  In the MSW
formalism, for example, $Re(n)$ is of the order of $G_F$, whereas
$Im(n)$ is of the order of $G_F^2q^2$.  In a CP symmetric plasma
such as the early universe, however, the order $G_F$ of $Re(n)$ cancels.
As we have already learned,  the leading terms of $Re(n)$ are
 now only of the
order of $G_Fq^2/M_W^2$.   Since $q^2\ll M_W^2$, in this situation
the imaginary  and the real part of
 index of refraction become interestingly comparable.

The simplest way
of calculating the imaginary part of index of refraction
is to employ the optical theorem
$$Im(n)={1\over 2p_0}\sum_jN_j\sigma_j,\eqno(7.1)$$
where $N_j$ is the density of the jth scatterer and
$\sigma_j$ is the corresponding total cross section.
Table 3 lists all the relevant scattering cross sections
normalized by\Ref\cross{The same results can also be found in
   the third article of Ref. 8.}
$$\sigma_0={G_F^2(p+p')^2\over 6\pi}.\eqno(7.2)$$
The result given by neutrino-photon scattering is completely negligible.

\bigskip

\chapter{Summary}

In this paper we have systematically evaluated the standard electroweak
interaction contributions to the neutrino index of refraction
in the early universe.  Of cosmological interest, we have
concentrated on the period when the temperature
of the universe was of the order of a few $MeV$, and the scatterers
were  photons, electrons, positrons and three types
of neutrinos and anti-neutrinos.

 Assuming CP invariance,
the number of particles and anti-particles is equal, and as a result
the lowest order coherent forward scattering amplitudes completely cancel.
Hence, the leading nonvanishing result obtained by the expansion
of the gauge boson propagator is very small.  In the absence
of leptonic family mixing,  which happens if
$\nu_e, \nu_{\mu}$ and $\nu_{\tau}$ are degenerate in mass,
 even such a tiny result vanishes
for scatterings between different families.  Although
the cancellation does not occur for  neutrino-photon scattering,
we find that its coherent forward scattering amplitude is zero
at one-loop level.

The smallness of tree level result motivates us to investigate
its radiative corrections. By employing the on-shell
renormalization scheme, we have found that depending on the specific
scattering process the leading log corrections are typically of
the order of $20\%$ to $10\%$ with the same sign.
Radiative corrections become the dominant contribution for scatterings
that involve different families if  mixings in the lepton sector
are very small.   Our results for neutrino index of refraction are
summarized in Table 2 and Table 4.
 The justification of our leading log approximation
relies on
 the hierarchy $M_{W,Z}^2\gg \vert q^2\vert,m_{\ell}^2$.

Numerically, these corrections already become significant.
They are about two orders of magnitude bigger than the original
expectation (Ref. \index) of the order of $\alpha/\pi$.
As summarized in Table 2 and Table 4, the radiative correction to the
neutrino index of refraction is about $+20\%$ and $+50\%$ for
$\nu_e$ and $\nu_{\mu,\tau}$, respectively.
This generates a sizable effect in locating
the exact boundaries from the nucleosynthesis
constraint for the aforementioned $\nu_e-\nu_s$ oscillation.
A detailed discussion of the application of our results to cases
of cosmological and astrophysical interest will be presented in
a forthcoming paper.

Besides the numerical significance, we have proposed a theoretical
method to evaluate the
 neutrino index of refraction at finite temperature.
This method makes the lowest order calculation much simpler than
that derived from the finite temperature field theory, and more importantly
allows us to evaluate higher order corrections in a straightforward way.
The underlying physics also appears to be more intuitive and
transparent.

We also studied the question of the cancellation of infrared divergence
in a coherent process.  We have shown that such cancellations
indeed take place in all order in  renormalizable perturbation
theory.  As a consequence, the cancellation of infrared
divergences will not spoil the coherent condition.

\ack

We wish to thank Prof. A. Sirlin for a number valuable
discussions. This work was supported in part by
the U.S. Department of Energy, contract DE-ACO2-76-ERO-3071.
\endpage

\endpage
\centerline{APPENDIX}

In this appendix we give some details of our calculation for
the three- and four-point functions.  As explained in the text,
we have ignored all $t$-channel one-particle-reducible graphs.
In the remaining graphs we keep all terms of order $\alpha$ and
only leading log terms of order $\alpha (q^2/M_W^2)$.  The order
$\alpha$ terms, although they do not contribute to the neutrino
index of refraction, provide a check on the calculation.
Our calculation is carried out in the Feynman gauge.
Due to the cancellation of particle-particle and
particle-anti-particle scattering, graphs with scalar
lines can be omitted in the leading log approximation.
Consequently, only the leading log terms of
our result are gauge invariant.

{\bf{Three-point Functions}}

The three-point function correction to $\nu_ee\to \nu_ee$
given by $\delta M_{6e+6f}$ in Eq. (3.17) of the text is obtained by
evaluating the one-loop vertex diagrams shown in Fig. 12.
For convenience, we
parameterize it as a sum of individual diagram contributions
$${ig\over\sqrt 2}[\bar{u}_e(p')\gamma_{\alpha}Lu_{\nu_e}(p)]
{\alpha\over 4\pi}
\sum_i\Gamma_{i},\eqno(A.1)$$
where
$$\Gamma_1(12a)=3\Bigl({2\over \epsilon}+\Gamma'(1)
+{5\over 6}-\ln{M_W^2\over 4\pi\mu_0^2}\Bigr)-{q^2\over M_W^2}
\Bigl(\ln{M_W^2\over m_e^2}+{7\over 6}\Bigr),\eqno(A.2)$$
$$\eqalign{
\Gamma_2(12b)= \Bigl({1\over 2s^2}-1\Bigr)\int^1_0dx\Bigl[&
6x\Bigl({2\over \epsilon}+\Gamma'(1)+{1\over 3}-\ln
{M_W^2\over 4\pi\mu_0^2}-\ln{1-xs^2\over c^2}-{1\over c^2}\ln
{c^2+xs^2\over x}\Bigr)\cr
&-{q^2\over M_Z^2}\Bigl({(x-1)(8x-2)\over c^2}+{x(8x-2s^2)\over c^4}
\ln{c^2+xs^2\over x}\Bigr)\Bigr],\cr}\eqno(A.3)$$
$$\Gamma_3(12c)={\Gamma_2(12b)\over (1-2s^2)},\eqno(A.4)$$
$$\eqalign{
\Gamma_4(12d)={1\over c^2}\Bigl(1-{1\over 2s^2}\Bigr)\int^1_0dx
\Bigl[&x\Bigl({2\over \epsilon}+\Gamma'(1)-1-\ln{xM_Z^2+(1-x)^2m_e^2
\over 4\pi\mu_0^2}\Bigr)\cr
&-{q^2\over M_Z^2}\Bigl(2x(x-1)\ln{xM_Z^2+(1-x)^2m_e^2\over (1-x)m_e^2
+x(x-1)q^2}-x^2\Bigr)\Bigr].\cr}\eqno(A.5)$$

Similarly,  for the
$\nu\nu\to\nu\nu$ scattering (Fig. 13), we have
$${ig\over 2c}[\bar{u}_{\nu}(p')\gamma_{\alpha}Lu_{\nu}(p)]
{\alpha\over 4\pi}\sum_i\Gamma_i',\eqno(A.6)$$
where
$$\Gamma'_1(13a)={1\over 4s^2c^2}\Bigl[{2\over\epsilon}+\Gamma'(1)
-{1\over 2}-\ln{M_Z^2\over 4\pi\mu_0^2}+{q^2\over 9M_Z^2}
+{4q^2\over M_Z^2}\int^1_0dxx(x-1)\ln{x(x-1)q^2+m_{\nu}^2\over M_Z^2}
\Bigr],\eqno(A.7)$$
$$\eqalign{\Gamma_2'(13b)=\Bigl(1-{1\over 2s^2}\Bigr)\Bigl[
{2\over\epsilon}&+
\Gamma'(1)-{1\over 2}-\ln{M_W^2\over 4\pi\mu_0^2}
+{5q^2\over 18 M_W^2}\cr
&+{4q^2\over M_W^2}
\int^1_0dxx(x-1)\ln{x(x-1)q^2+m_{\ell}^2\over M_W^2}
\Bigr],\cr}\eqno(A.8)$$
$$\Gamma_3'(13c)={c^2\over s^2}\Bigl[3\Bigl({2\over \epsilon}
+\Gamma'(1)-{1\over 6}-\ln{M_W^2\over 4\pi\mu_0^2}\Bigr)
+{q^2\over 3M_W^2}
\Bigr].\eqno(A.9)$$

{\bf{Box Diagrams}}

First, the photon box (Fig. 4c) result is
$$\eqalign{\delta M_{Box}^{QED}=&{-ig^2\over 2M_W^2}{\alpha\over 4\pi }
[\bar{u}_e(p')\gamma_{\alpha}Lu_e(p')]
[\bar{u}_{\nu_e}(p)\gamma^{\alpha}Lu_{\nu_e}(p)]\cr
&\times\Bigl[\Bigl(1+{q^2\over M_W^2}\Bigr)\Bigl(
2\ln {\lambda^2\over m_e^2}+\ln {M_W^2\over m_e^2}+{7\over 2}\Bigr)
-{q^2\over M_W^2}\Bigl({2\over 3}\ln{M_W^2\over m_e^2}
-{19\over 9}\Bigr)\Bigr]
\cr
&{-ig^2\over 2M_W^2}{\alpha\over 4\pi }
[\bar{u}_e(p')\gamma_{\alpha}Ru_e(p')]
[\bar{u}_{\nu_e}(p)\gamma^{\alpha}Lu_{\nu_e}(p)]
\Bigl(1+{1\over 3}{q^2\over M_W^2}\Bigr),\cr}\eqno(A.10)$$
where $\lambda$ is a ficticious photon mass.

In calculating the remaining weak interaction box diagrams, we
frequently encounter a scalar integral
$$F_{\alpha\beta}(m,m_{\ell},q^2)\equiv
\int {d^4K\over (2\pi)^4}{(K-q)_{\alpha}K_{\beta}\over
(K^2-m^2)[(K-q)^2-m_{\ell}^2]
[(K-p)^2-M_W^2][(K-p')^2-M_Z^2]}.\eqno(A.11a)$$
An explicit calculation shows
$$\eqalign{&F_{\alpha\beta}(m,m_{\ell},q^2)=\cr
&{-i\over 16\pi^2M_W^2}
\Bigl({c^2\over s^2}\Bigr)
\Bigl\{-{\ln c^2\over 4}g_{\alpha\beta}+{g_{\alpha\beta}\over 2}
\Bigl[K_0(m,m_{\ell}, M_W,q^2)-K_0(m,m_{\ell},M_Z,q^2)\Bigr]\cr
&+q_{\alpha}q_{\beta}\Bigl[{1\over M_W^2}K_1(m,m_{\ell},M_W,q^2)-
{1\over M_Z^2}K_1(m,m_{\ell},M_Z,q^2)\Bigr]\Bigr\}\cr
&+...,\cr}\eqno(A.11b)$$
where the ellipses represent the  terms which do not have a
large log, and
$$\eqalign{&K_0(m,m_{\ell}, M, q^2)\cr
&\equiv \int^1_0dx
{m^2+x(m_{\ell}^2-m^2)+x(x-1)q^2\over M^2}
\ln {m^2+x(m_{\ell}^2-m^2)+x(x-1)q^2\over M^2},\cr
&K_1(m,m_{\ell},M,q^2)\equiv \int^1_0dxx(x-1)\ln
{m^2+x(m_{\ell}^2-m^2)+x(x-1)q^2\over M^2}.\cr}\eqno(A.12a)$$
They are related to the functions $I_{1,2}$ (Eq. (3.1)) by
$$\eqalign{
&K_0(m,m_{\ell},M,q^2)=I_1(m,m_{\ell},M,q^2)+I_2(m,m_{\ell},M,q^2),\cr
&K_1(m,m_{\ell},M,q^2)={M^2\over q^2}I_1(m,m_{\ell},M,q^2).\cr}
\eqno(A.12b)$$

{}From the analyticity of scalar integrals\rlap,\Ref\integral{
      G. 't Hooft and M. Veltman,  Nucl. Phys. B153, 365 (1979).}
one can show that for our case only those graphs which have an
$s$ and/or
$u$-channel light intermediate state have a large log.  For the
$\nu_ee\to\nu_ee$ scattering (Fig. 7), we find
$$\eqalign{
\delta M(7a)={-ig^2\over 2M_W^2}{\alpha\over 4\pi}&[\bar{u}_e(p')
\gamma_{\alpha}Lu_{\nu_e}(p)][\bar{u}_{\nu_e}(p)\gamma^{\alpha}Lu_e(p')]
\cr
&\times \Bigl(1-{1\over 2s^2}\Bigr)^2\Bigl[\ln c^2+
{2q^2\over 3M_W^2}{c^2\over s^2}\Bigl(1+{1+c^2\over 2 s^2}\ln c^2\Bigr)
\Bigr],\cr}\eqno(A.13)$$
$$
\delta M(7b)= {\delta M(7a)\over (2s^2-1)^2},\eqno(A.14)$$
$$\eqalign{\delta M(7c)=\delta M(7d)=&{ig^4\over 32\pi^2 M_Z^2c^2s^2}
(2s^2-1)[\bar{u}_e(p')\gamma_{\alpha}Lu_{\nu_e}(p)][\bar{u}_{\nu_e}(p)
\gamma^{\alpha}Lu_e(p')]\cr
&\times\Bigl\{\ln c^2-2\Bigl[K_0(0,m_e,M_W,-q^2)-K_0(0,m_e,M_Z,-q^2)
\Bigr]\cr
& +q^2\Bigl[{1\over M_W^2}K_1(0,m_e,M_W,-q^2)
-{1\over M_Z^2}K_1(0,m_e,M_Z,-q^2)\Bigr]\Bigr\},\cr}\eqno(A.15)$$
$$\eqalign{\delta M(7e)={ig^4\over 64\pi^2M_Z^2c^4}&[\bar{u}_e(p')
\gamma_{\alpha}(4s^4+(1-4s^2)L)u_e(p')][\bar{u}_{\nu_e}(p)\gamma^{\alpha}
Lu_{\nu_e}(p)] \cr
&\times \Bigl[1+2K_0(0,m_e,M_Z,-q^2)-{q^2\over M_Z^2}
K_1(0,m_e,M_Z,-q^2)\Bigr],\cr}\eqno(A.16)$$
$$\eqalign{\delta M(7f)&={-ig^4\over 64\pi^2M_Z^2c^4}
[\bar{u}_e(p')\gamma_{\alpha}(4s^4+(1-4s^2)L)u_e(p')]
[\bar{u}_{\nu_e}(p)\gamma^{\alpha}Lu_{\nu_e}(p)]\cr &\ \ \ \ \ \ \
\times\Bigl[
{1\over 4}+{1\over 2}K_0(0,m_e,M_Z, q^2)\Bigr] \cr
&-{ig^4\over 64\pi^2M_Z^4c^4}[\bar{u}_e(p')\gamma_{\alpha}
(4s^4+(1-4s^2)L)u_e(p')][\bar{u}_{\nu_e}(p)\gamma_{\beta}
Lu_{\nu_e}(p)]\cr &\ \ \ \ \ \ \
\times q^{\alpha}q^{\beta}K_1(0,m_e,M_Z,q^2),\cr}\eqno(A.17)$$
$$\eqalign{\delta M(7g)&={-ig^4\over 16\pi^2M_W^2}
[\bar{u}_e(p')\gamma_{\alpha}Lu_e(p')]
[\bar{u}_{\nu_e}(p)\gamma^{\alpha}Lu_{\nu_e}(p)]\Bigl[{1\over 4}
+{1\over 2}K_0(0,m_e,M_W,q^2)\Bigr]\cr
&-{ig^4\over 16\pi^2M_W^4}[\bar{u}_e(p')\gamma_{\alpha}Lu_e(p')]
[\bar{u}_{\nu_e}(p)\gamma_{\beta}Lu_{\nu_e}(p)]
q^{\alpha}q^{\beta}K_1(0,m_e,M_W,q^2),\cr}\eqno(A.18)$$
$$\eqalign{\delta M(7h)={ig^4\over 16\pi^2M_W^2}&
[\bar{u}_e(p')\gamma_{\alpha}Lu_e(p')]
[\bar{u}_{\nu_e}(p)\gamma^{\alpha}Lu_{\nu_e}(p)]\cr
&\times
\Bigl[1+2K_0(m_e,0,M_W,-q^2)-{q^2\over M_W^2}
K_1(m_e,0,M_W,-q^2)\Bigr].\cr}
\eqno(A.19)$$

For the scattering $\nu\nu\to \nu\nu$ (Fig. 11), we find
$$\eqalign{\delta M(11a)=&{ig^4\over 64\pi^2M_Z^2c^4}
[\bar{u}_{\nu}(p)\gamma_{\alpha}Lu_{\nu}(p)]
[\bar{u}_{\nu}(p')\gamma^{\alpha}Lu_{\nu}(p')]\cr
&\times\Bigl[1+
2K_0(0,0,M_Z,-q^2)-{q^2\over M_Z^2}K_1(0,0,M_Z,-q^2)\Bigr],\cr}
\eqno(A.20)$$
$$\eqalign{
\delta M(11b)={-ig^4\over 64\pi^2M_Z^2c^4}&
[\bar{u}_{\nu}(p)\gamma_{\alpha}Lu_{\nu}(p)]
[\bar{u}_{\nu}(p')\gamma^{\alpha}Lu_{\nu}(p')]\cr
&\times\Bigl[{1\over 4}+{1\over 2}K_0(0,0,M_Z,q^2)\Bigr]\cr
{-ig^4\over 64\pi^2M_Z^4c^4}&
[\bar{u}_{\nu}(p)\gamma_{\alpha}Lu_{\nu}(p)]
[\bar{u}_{\nu}(p')\gamma_{\beta}Lu_{\nu}(p')]\cr
&\times q^{\alpha}q^{\beta}K_1(0,0,M_Z,q^2),\cr}\eqno(A.21)$$
$$\eqalign{\delta M(11c)={-ig^4\over 64\pi^2M_W^2}
&[\bar{u}_{\nu}(p)\gamma_{\alpha}Lu_{\nu}(p)]
[\bar{u}_{\nu}(p')\gamma^{\alpha}Lu_{\nu}(p')]\cr
&\times \Bigl[1+2K_0(m_{\ell},m_{\ell},M_W, q^2)\Bigr]\cr
-{ig^4\over 16\pi^2M_W^4}&
[\bar{u}_{\nu}(p)\gamma_{\alpha}Lu_{\nu}(p)]
[\bar{u}_{\nu}(p')\gamma_{\beta}Lu_{\nu}(p')]\cr
&\times q^{\alpha}q^{\beta}K_1(m_{\ell}
,m_{\ell},M_W,q^2),\cr}\eqno(A.22)$$
$$\eqalign{\delta M(11d)={-ig^4\over 64\pi^2M_Z^2c^4}&
[\bar{u}_{\nu}(p)\gamma_{\alpha}Lu_{\nu}(p)]
[\bar{u}_{\nu}(p')\gamma^{\alpha}Lu_{\nu}(p')]\cr
&\times \Bigl[{1\over 4}+{1\over 2}K_0(0,0, M_Z,0)\Bigr],\cr}
\eqno(A.23)$$
$$\eqalign{\delta M(11e)={ig^4\over 64\pi^2M_Z^2c^4}&
[\bar{u}_{\nu}(p)\gamma_{\alpha}Lu_{\nu}(p)]
[\bar{u}_{\nu}(p')\gamma^{\alpha}Lu_{\nu}(p')]\cr
&\times \Bigl[1+2K_0(0,0,M_Z,-q^2)-{q^2\over M_Z^2}
K_1(0,0,M_Z,-q^2)\Bigr],\cr}\eqno(A.24)$$
$$\eqalign{\delta M(11f)={-ig^4\over 64\pi^2M_W^2}&
[\bar{u}_{\nu}(p)\gamma_{\alpha}Lu_{\nu}(p)]
[\bar{u}_{\nu}(p')\gamma^{\alpha}Lu_{\nu}(p')]\cr
&\times \Bigl[1+2K_0(m_{\ell},m_{\ell},M_W,0)\Bigr].\cr}
\eqno(A.25)$$
\endpage

\refout
\endpage

\centerline{FIGURE CAPTIONS}

Fig.1. Tree level graphs for $\nu_e e\to\nu_e e$ forward scattering.
The coherent condition only allows the $W$ in the $u$-channel (1a) to
carry a non-zero momentum.

Fig.2. Tree level graphs for $\nu_e\bar{e}\to\nu_e\bar{e}$ forward
scattering.  Again, only the $W$ in the $s$-channel (2a) is allowed
to carry a non-zero momentum.

Fig. 3. Feynman graphs for an electron forward scattering
with an infrared finite but otherwise arbitrary potential (the blob)
and its QED corrections.  The ellipses represent other
graphs which are infrared finite.

Fig. 4.  One-loop QED corrections to the $u$-channel neutrino-electron
forward scattering. The photon propagator in (4a) and (4b) is the
massless photon propagator (the second term in Eq. (3.10)).
The box diagram (4c) is evaluated with the full photon propagator.

Fig. 5. One-loop self-energy and counterterm contributions
to the $u$-channel neutrino-electron forward scattering.

Fig. 6.  Vertex and wave function renormalization
 corrections to the $u$-channel neutrino-electron
scattering.  The photon propagators  in (6a) and (6b) are the massive
photon propagators (the first term in Eq. (3.10)).  The detail
of the blob is shown in Fig.12.

Fig. 7. Box diagram contributions to $\nu_ee\to\nu_ee$ forward
scattering.

Fig. 8.  Tree level diagrams for $\nu\nu\to\nu\nu$
and $\nu\bar{\nu}\to\nu\bar{\nu}$  forward scattering.

Fig. 9.  One-loop self-energy and counterterm contributions
to $\nu\nu\to\nu\nu$ forward scattering.

Fig. 10. Vertex corrections to the exchange channel
in $\nu\nu\to\nu\nu$ scattering.  The detail of the blob
is shown in Fig. 13.  The ellipses represent the wave function
renormalization diagrams.

Fig. 11.  Box diagram contributions to $\nu\nu\to\nu\nu$ forward
scattering.

Fig. 12.  One-loop vertex diagrams for $\nu_ee\to\nu_ee$
forward scattering.  Here the photon contribution is evaluated
with the full propagator.

Fig. 13. One-loop vertex diagrams for $\nu\nu\to\nu\nu$ forward
scattering.
\endpage
\centerline{TABLE CAPTIONS}

Table 1.  Radiative corrections to the neutrino index of refraction,
assuming the medium is CP symmetric.  Here $\vert q\vert ^2$
varies from $1\ MeV^2$ to $400\ MeV^2$.  The definition of
the various corrections is given in the text.    A $1\%$ correstion
should be added to the diagonal $\nu\nu\to\nu\nu$ scattering if the
top quark mass is within $150\ GeV$ to $200\ GeV$.  Results for
scattering between different families in the small mixing limit
 are given in Eqs. (5.3)
and (5.5) of the text, and are summarized in Table 2.

Table 2.  Summary of leading log results of radiative corrections.
The interaction process also includes the corresponding scattering
with the anti-scatterers.

Table 3.  The total scattering cross section for neutrinos
with the various scatterers in the early universe.  The result
is calculated in the center of mass systerm and
 normalized by $\sigma_0$ (Eq. (7.2)).   The plus-sign refers to
$i=e$ and the minus-sign corresponds to $i=\mu,\tau.$  Also, $i\ne j$.

Table 4. Neutrino index of refraction in the early universe.  Here,
$n_0$ is given by
$n_0\equiv -(2\sqrt 2/M_W^2)G_{\mu}N_{\gamma}\langle p_0'\rangle $.
The densities of the scatterers are normalized in terms of
the photon density according to Eq. (1.1).  The various
radiative correction results, $\delta_{ij}$, are summarized
in Table 2.

\endpage
\input tables
\begintable
\tstrut\  Sourses|$\nu_ee(\bar{e})\to \nu_ee(\bar{e})$
|$\nu_e\nu_e(\bar{\nu}_e)\to \nu_e\nu_e(\bar{\nu}_e)
$|$\nu_{\mu,\tau}\nu_{\mu,\tau}(\bar{\nu}_{\mu,\tau})
\to\nu_{\mu,\tau}\nu_{\mu,\tau}(\bar{\nu}_{\mu,\tau})$\crthick
hadronic self-energy:$5\%$:$5\%$:$5\%$\nr
leptonic self-energy:$4\%\ $ to$\ 3\%$:$7\%\ $to$\ 5\%$:
$7\%\ $to $\ 5\%$\nr
bosonic self-energy:negligible:negligible:negligible\nr
vertex+box:$\le 2\%$:$13\%\ $ to $\ 10\%$:$12\%\ $to $\ 10\%$\nr
$QED$:$-1\%$:$\ $:$\ $\cr
total:$10\%\ $ to $\ 7\%$:$25\%\ $ to $\ 20\%$:
$24\%\ $ to $\ 20\%$\endtable

$$ $$
\centerline{Table 1}
\endpage

\begintable
\tstrut\ correction: interaction: result $(\vert q\vert=1\ MeV)$:
result $(\vert q\vert =20\ MeV)$\crthick
$\delta_{\nu_e\nu_e}$:$\nu_e\nu_e\to
                       \nu_e\nu_e$:$25\%$:$20\%\ $\nr
$\delta_{\nu_{\mu}\nu_{\mu}}$:$\nu_{\mu}\nu_{\mu}\to\nu_{\mu}\nu_{\mu}$:
                                   $24\%$:$20\%\ $\nr
$\delta_{\nu_{\tau}\nu_{\tau}}$:$\nu_{\tau}\nu_{\tau}\to
                       \nu_{\tau}\nu_{\tau}$:$24\%$:$20\%\ $\nr
$\delta_{\nu_e\nu_{\mu}}$:$\nu_e\nu_{\mu}\to\nu_e\nu_{\mu}$:
                                   $7\%$:$6\%\ $\nr
$\delta_{\nu_e\nu_{\tau}}$:$\nu_e\nu_{\tau}\to\nu_e\nu_{\tau}$:$
                                    6\%$:$5\%\ $\nr
$\delta_{\nu_{\mu}\nu_{\tau}}$:$\nu_{\mu}\nu_{\tau}\to\nu_{\mu}
           \nu_{\tau}$:$6\%$:$5\%\ $\nr
$\delta_{\nu_ee}$:$\nu_ee\to\nu_ee$:$10\%$:$7\%\ $\nr
$\delta_{\nu_{\mu}e}$:$\nu_{\mu}e\to\nu_{\mu}e$:$11\%$:$10\%\ $\nr
$\delta_{\nu_{\tau}e}$:$\nu_{\tau}e\to\nu_{\tau}$:$7\%$:$6\%\ $
\endtable

$$ $$
\centerline{Table 2}
\endpage

\begintable
\tstrut\ interaction-channel: cross-section\crthick
$\nu_i\bar{\nu}_i\leftrightarrow e\bar{e}$:$8s^4\pm 4s^2+1$\nr
$\nu_i\bar{\nu}_i\leftrightarrow \nu_j\bar{\nu}_j$:$1$\nr
$\nu_ie\leftrightarrow
\nu_ie$:$8s^4\pm 6s^2+{3\over 2}$\nr
$\nu_i\bar{e}\leftrightarrow \nu_i\bar{e}$:$8s^4\pm 2s^2+{1\over 2}$\nr
$\nu_i\nu_i\leftrightarrow \nu_i\nu_i$:$6$\nr
$\nu_i\nu_j\leftrightarrow \nu_i\nu_j$:$3$\nr
$\nu_i\bar{\nu}_i\leftrightarrow \nu_i\bar{\nu}_i$:$4$\nr
$\nu_i\bar{\nu}_j\leftrightarrow \nu_i\bar{\nu}_j$:$1$\nr
$\nu(\bar{\nu})\gamma\leftrightarrow \nu(\bar{\nu})\gamma$:
negligible\endtable

$$ $$
\centerline{Table 3}
\endpage

\begintable
\tstrut\ real part of index of refraction: results\crthick
$\ $:$\ $\nr
$(n-1)_{\nu_e}/n_0$:$2(1+\delta_{\nu_ee})+{1\over 2}[c^2\rho^{(\nu;\nu)}(
1+\delta_{\nu_e\nu_e})+\delta_{\nu_e\nu_{\mu}}+\delta_{\nu_e\nu_{\tau}}
]$\nr
$\ $:$\ $\nr
$(n-1)_{\nu_{\mu}}/n_0$:
$\delta_{\nu_{\mu}e}+{1\over 2}
[c^2\rho^{(\nu;\nu)}(1+\delta_{\nu_{\mu}\nu_{\mu}})
+\delta_{\nu_{\mu}\nu_e}+\delta_{\nu_{\mu}\nu_{\tau}}]$\nr
$\ $:$\ $\nr
$(n-1)_{\nu_{\tau}}/n_0$:
$\delta_{\nu_{\tau}e}+{1\over 2}
[c^2\rho^{(\nu;\nu)}
(1+\delta_{\nu_{\tau}\nu_{\tau}})+\delta_{\nu_{\tau}\nu_e}
+\delta_{\nu_{\tau}\nu_{\mu}}]$\nr $\ $:$\ $\endtable

$$ $$
\centerline{Table 4}
\endpage

\bye